\documentclass[numbers,webpdf,imamat]{for-arxiv-template}

\usepackage{stmaryrd}
\usepackage{xcolor,soul}
\usepackage{diagbox}
\usepackage{multirow}
\usepackage{enumerate}
\usepackage{tensor}

\usepackage{newtxmath} 

\newcommand{\wrt}{w.\,r.\,t.}
\newcommand{\eg}{e.\,g.,}
\newcommand{\ie}{i.\,e.,}

\newcommand{\resp}{resp.}
\newcommand{\cf}{cf.}

\newcommand{\formComma}{\,\text{,}}
\newcommand{\formPeriod}{\,\text{.}}

\newcommand{\surf}{\mathcal{S}}
\newcommand{\U}{\mathcal{U}}
\newcommand{\R}{\mathbb{R}}

\newcommand{\tangent}[2][]{\tensor{\operatorname{T}\!}{#1}#2}
\newcommand{\tangentS}[1][]{\tangent[#1]{\surf}}
\newcommand{\tangentR}[1][]{\tangent[#1]{\R^3\vert_{\surf}}}

\newcommand{\tangentStar}[2][]{\tensor*{\operatorname{T}\!}{#1}#2}

\newcommand{\hil}{\operatorname{L}^{\!2}}
\newcommand{\hilbert}[1]{\hil(#1)}
\newcommand{\hilbertT}[2][]{\hilbert{\tangent[#1]{#2}}}
\newcommand{\hilbertS}[1][]{\hilbert{\tangentS[#1]}}
\newcommand{\hilbertR}[1][]{\hilbert{\tangentR[#1]}}
\newcommand{\inner}[3][0]{\left\langle #3 \right\rangle_{#2}} 
\newcommand{\innerH}[3][0]{\inner[#1]{\hilbert{#2}}{#3}}

\newcommand{\hilgrad}{\nabla_{\hil}}

\newcommand{\gb}{\boldsymbol{g}}
\newcommand{\Wb}{\boldsymbol{W}}
\newcommand{\wb}{\boldsymbol{w}}
\newcommand{\wnor}{w_{\bot}}
\newcommand{\qb}{\boldsymbol{q}}
\newcommand{\Qb}{\boldsymbol{Q}}
\newcommand{\rb}{\boldsymbol{r}}
\newcommand{\Rb}{\boldsymbol{R}}
\newcommand{\eb}{\boldsymbol{e}}
\newcommand{\bb}{\boldsymbol{b}}
\newcommand{\Gb}{\boldsymbol{G}}
\newcommand{\Sb}{\boldsymbol{S}}
\newcommand{\Ab}{\boldsymbol{A}}
\newcommand{\Eb}{\boldsymbol{E}}

\newcommand{\Psib}{\boldsymbol{\Psi}}
\newcommand{\Phib}{\boldsymbol{\Phi}}
\newcommand{\Vb}{\boldsymbol{V}}
\newcommand{\vb}{\boldsymbol{v}}
\newcommand{\vnor}{v_{\bot}}

\newcommand{\ub}{\boldsymbol{u}}
\newcommand{\sigmab}{\boldsymbol{\sigma}}
\newcommand{\etab}{\boldsymbol{\eta}}
\newcommand{\nullb}{\boldsymbol{0}}

\newcommand{\Qcal}{\mathcal{Q}}

\newcommand{\mfrak}{\mathfrak{m}}
\newcommand{\ofrak}{\mathfrak{o}}

\newcommand{\gstress}{\sigmab}
\newcommand{\tgstress}{\widetilde{\gstress}}
\newcommand{\hgstress}{\widehat{\gstress}}

\newcommand{\paraC}{X}
\newcommand{\para}{\boldsymbol{\paraC}}
\newcommand{\normalC}{\nu}
\newcommand{\normal}{\boldsymbol{\normalC}}
\newcommand{\shopC}{I\!I}
\newcommand{\shop}{\boldsymbol{\shopC}}

\newcommand{\meanc}{\mathcal{H}}
\newcommand{\gaussc}{\mathcal{K}}

\newcommand{\gradW}[1][\Wb]{\boldsymbol{G}[\para,#1]}

\newcommand{\eps}{\varepsilon}

\newcommand{\curlyd}{\eth}

\newcommand{\jau}{\mathcal{J}}
\newcommand{\rot}{\operatorname{rot}}
\renewcommand{\div}{\operatorname{div}}
\newcommand{\Tr}{\operatorname{Tr}}
\newcommand{\gauge}[2][]{\tensor*{\curlyd}{_{#2}^{#1}}}
\newcommand{\gaugeS}[1][]{\gauge[\surf]{#1}}
\newcommand{\gaugeJ}[1][]{\gauge[\jau]{#1}}
\newcommand{\gaugeU}[1][]{\gauge[\sharp]{#1}}
\newcommand{\gaugeL}[1][]{\gauge[\flat]{#1}}
\newcommand{\gaugeN}[1][]{\gauge[\natural]{#1}}
\newcommand{\gaugeUU}[1][]{\gauge[\sharp\sharp]{#1}}
\newcommand{\gaugeLL}[1][]{\gauge[\flat\flat]{#1}}
\newcommand{\gaugeUL}[1][]{\gauge[\sharp\flat]{#1}}
\newcommand{\gaugeLU}[1][]{\gauge[\flat\sharp]{#1}}

\newcommand{\timeD}{\mathfrak{D_{t}}}
\newcommand{\timeLu}{\mathfrak{L}^{\sharp}}
\newcommand{\timeLl}{\mathfrak{L}^{\flat}}
\newcommand{\timeJ}{\mathfrak{J}}

\newcommand{\timeLuu}{\mathfrak{L}^{\sharp\sharp}}
\newcommand{\timeLll}{\mathfrak{L}^{\flat\flat}}
\newcommand{\timeLul}{\mathfrak{L}^{\sharp\flat}}
\newcommand{\timeLlu}{\mathfrak{L}^{\flat\sharp}}

\newcommand{\del}{\operatorname{\delta}\!}
\newcommand{\adj}{\dagger}
\newcommand{\dadj}{\ddagger}

\newcommand{\proj}{\operatorname{\Pi}}
\newcommand{\projb}{\boldsymbol{\proj}}
\newcommand{\Id}{\operatorname{Id}}

\newcommand{\landau}{\mathcal{O}}

\newcommand{\energy}{\mathfrak{U}}
\newcommand{\tenergy}{\widetilde{\energy}}
\newcommand{\henergy}{\widehat{\energy}}
\newcommand{\rayleighian}{\mathfrak{R}}

\definecolor{lighter_purple_mathematica}{rgb}{0.6666666666,0.33333333333,0.666666666666}

\theoremstyle{thmstyleone}%
\theoremstyle{thmstyletwo}%
\newtheorem{remark}{Remark}%

\theoremstyle{thmstylethree}%
\newtheorem{definition}{Definition}%

\begin{document}

\DOI{DOI HERE}
\copyrightyear{2021}
\vol{00}
\pubyear{2023}
\access{Advance Access Publication Date: Day Month Year}
\appnotes{Paper}
\copyrightstatement{Published by Oxford University Press on behalf of the Institute of Mathematics and its Applications. All rights reserved.}
\firstpage{1}

\title[Tangential Tensor Fields on Deformable Surfaces]{Tangential Tensor Fields on Deformable Surfaces -- How to Derive Consistent $L^2$-Gradient Flows}

\author{Ingo Nitschke*
\address{\orgdiv{Institute of Scientific Computing}, \orgname{Technische Universität Dresden}, \orgaddress{Dresden, \postcode{01062}, \country{Germany}}}}
\author{Souhayl Sadik
\address{\orgdiv{Department of Mechanical and Production Engineering}, \orgname{Aarhus University}, \orgaddress{Aarhus, \postcode{8000}, \country{Denmark}}}}
\author{Axel Voigt
\address{\orgdiv{Institute of Scientific Computing}, \orgname{Technische Universität Dresden}, \orgaddress{Dresden, \postcode{01062}, \country{Germany}}}
\address{\orgdiv{Dresden Center for Computational Materials Science (DCMS)}, \orgname{Technische Universität Dresden}, \orgaddress{Dresden, \postcode{01062}, \country{Germany}}}
\address{\orgname{Center for Systems Biology Dresden (CSBD)}, \orgaddress{\street{Pfotenhauerstr. 108}, Dresden, \postcode{01307}, \country{Germany}}}
}

\authormark{I. Nitschke et al.}

\corresp[*]{Corresponding author: \href{email:ingo.nitschke@tu-dresden.de}{ingo.nitschke@tu-dresden.de}}

\received{Date}{0}{Year}
\revised{Date}{0}{Year}
\accepted{Date}{0}{Year}

\abstract{We consider gradient flows of surface energies which depend on the surface by a parameterization and on a tangential tensor field. The flow allows for dissipation by evolving the parameterization and the tensor field simultaneously. This requires the choice of a notation for independence. We introduce different gauges of surface independence and show their consequences for the evolution. In order to guarantee a decrease in energy, the gauge of surface independence and the time derivative have to be chosen consistently. We demonstrate the results for a surface Frank-Oseen-Hilfrich energy. \\}

\maketitle

\section{Introduction}

Gradient flows are evolutionary systems that decrease energy through a dissipation mechanism. Such systems are common in physics and widely explored in mathematics \cite{Ambrosio_book_2008}. In this work, we consider classical Hilbert spaces, the $L^2$-norm as a dissipation mechanism, and surface energies $ \energy[\para,\qb[\para]] $ that depend on a surface $\surf$ through a parameterization $\para$ and a tangential n-tensor field $ \qb[\para]\in\tangentS[^n]$\textemdash see section \ref{sec:notations} for definitions. The tangential n-tensor field $\qb[\para]$ is defined with respect to the frame in its image space, which introduces a dependency on the parameterization $\para$. 
$L^2$-gradient flows of such surface energies, which allow for dissipation by evolving $\para$ and $\qb[\para],$ simultaneously, are given by:

\begin{equation} \label{eq:1}
\left( \Vb[\para], \timeD\qb[\para] \right) = -\lambda \hilgrad\energy,
\end{equation}

Here, $\Vb$ denotes the velocity of the surface $\surf$, and $\timeD\qb[\para]$ denotes an observer-invariant instantaneous time derivative of the n-tensor field $\qb[\para]$, see \cite{Nitschke_2022}. This work is concerned with providing a precise mathematical definition for the $L^2$-gradient on the right-hand side of \eqref{eq:1}. The delicate issue is the requirement of mutual independence of $\para$ and $\qb[\para]$:\footnote{Without any dependency on $\qb[\para]$, \eg\ for $\energy = \energy[\para]$, the $L^2$-gradient flow \eqref{eq:1} comprises the first variation of $\energy$ \wrt\ $\para$, which could be seen as the shape derivative in its strong form, as discussed in \cite{Henrot_2018,Ito_2006,Delfour_1991}.} What does this actually mean? When is a tensor field independent of its underlying spatial domain? We investigate these issues for variations of the first kind and reveal consequences for such $ \hil $-gradient flows, \eg\ the requirement to choose the notion of independence consistent with the time derivative. Furthermore, we provide physical cues to choose an appropriate notion of independence or a related time derivative. While this choice might be intuitive for elastic surfaces \cite{Gurtin_ARMA_1975,steigmann2007thin,Yavarietal_JNS_2016,sadik2016shells,Lewicka_book_2022,pezzulla2022geometrically} and fluidic surfaces \cite{scriven1960dynamics,Arroyoetal_PRE_2009,Reutheretal_MMS_2015,Kobaetal_QAM_2017,jankuhn2018,Miura_QAM_2018,Reutheretal_MMS_2018,Kobaetal_QAM_2018,torres2019modelling,reuther2020numerical}, this changes for viscoelastic surfaces \cite{Snoeijer_PRSA_2020,deKinkelder_JCP_2021, sadik2023nonlinear} and surface liquid crystals \cite{Reuteretal_PRSA_2019_Preprint,Reuteretal_PRSA_2020, mihai2020plate, ozenda2020blend, singh2022ribbon,bartels2022lce}. Crucial examples where the last two are of relevance are biological surfaces, \eg\ the actomyosin cortex, which drives cell surface deformations \cite{reymann2016cortical,naganathan2018morphogenetic}, or on a larger scale, epithelia tissue, which deforms during morphogenesis \cite{Maroudas-Sacks_NP_2021}. Existing models for such applications are still in their infancy \cite{Mietke_PNAS_2019,deKinkelder_JCP_2021,Maroudas-Sacks_NP_2021,Hoffmann_SA_2022,SalbreuxJuelicherProstCallan-Jones__2022,Morris_2022}. However, they all share the feature of choosing the notions of independence and time derivative in an ad hoc fashion. As these systems are by definition out of equilibrium, their dynamics matter. We will demonstrate the dependency of the dynamics on the notion of independence. Any quantitative study of such systems, therefore, requires dealing with these issues.

The paper is structured as follows: In section \ref{sec:notations}, we introduce linear deformed surfaces and some necessary mathematical tools, which cover independence of coordinates and leave the surface as a degree of freedom. In section \ref{sec:goi_main}, we consider deformation derivatives for tensor fields, which leads us to the concept of the gauge of surface independence. Treating scalar fields in subsection \ref{scalar_fields} is merely a warm-up exercise to make the reader familiar with the concept of the deformation derivative and surface independence. Tangential vector- and tensor-fields are treated in subsection \ref{sec:vector_fields} and \ref{sec:tangential_tensor_fields}. For instance, we see that the variation of an energy depending on the contravariant proxy of a tensor field\textemdash which is assumed to be independent of the surface\textemdash leads to a different result than the same variation for the same energy but with a covariant proxy assumed to be independent. It cannot be due to the covariant or contravariant frame, since the calculus is invariant \wrt\ frames, \ie\ musical isomorphisms. It must be, then, because of the assumption of independence instead. We elaborate on this discrepancy for tangential vector fields in subsection \ref{sec:vector_fields}, introduce four different gauges of surface independence, and relate them mutually. That includes some useful identities for calculating surface variations. We also show the influence on the $ \hil $-gradient out of an equilibrium state, which, in turn, does have an impact on the $ \hil $-gradient flow. In section \ref{sec:tangential_tensor_fields}, we extend our results to $ 2 $-tensor fields, thus providing the reader with a roadmap for further extensions to $ n $-tensor fields. We illustrate the difference between the various gauges of surface independence in an example in section \ref{sec:Examples}, where the $ \hil $-gradient flow for a special case of a surface Frank-Oseen-Helfrich energy is solved numerically under all gauges of surface independence. Finally, we draw conclusions in section \ref{sec:Conclusions}.

\section{Notations and Mathematical Preliminaries} \label{sec:notations}

\subsection{Surfaces and Linear Deformations}
For our purposes, a surface is a 2-dimensional Riemannian manifold $ (\surf,\gb) $ smoothly embededed in the 3-dimensional Euclidean space $ \R^3 $ by a smooth parameterization
\begin{align}
	\para=\begin{bmatrix}
                \paraC^1\\ \paraC^2 \\ \paraC^3
            \end{bmatrix}:\quad \U \rightarrow \R^3:\quad (y^1,y^2) \mapsto \para(y^1,y^2)\in\surf \formComma
\label{eq:parameterization}
\end{align}
where $ \U\subset\R^2 $ is an open set. For simplicity, we assume that $\surf$ can be realised by a single parameterization $\para$, \ie\ $ \para(\U)=\surf $. The ensuing results may be extended to the more general case by using an open covering of $\surf$. Since $\para$ determines $\surf$, the parameterization may be regarded as an independant variable describing the dependance on the surface itself. This has the advantage that $\para$ is defined in the embedding space, which bears vector space structure where one may work with vector calculus.

\begin{figure}
\centering
\includegraphics[width=0.99\linewidth]{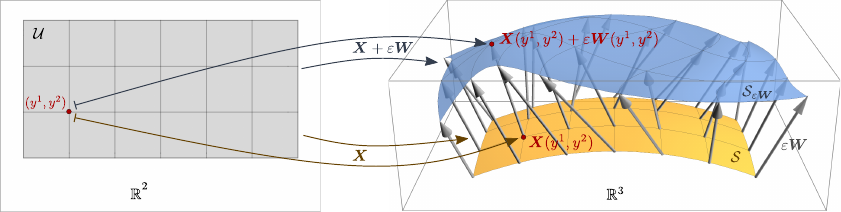}       
\caption{Schematic of parameterization and deformation: $ \U\subset\R^2 $ is the domain \textit{(left)} of the parameterization 
        $ \para $, \resp\ $ \para+\eps\Wb $, which maps to the surface $ \surf\subset\R^3 $ \textit{(lower right)}, \resp\ $ \surf_{\eps\Wb}\subset\R^3 $ \textit{(upper right)}.
        The surface $ \surf $ is linearly deformed by $ \eps>0 $ along the direction of the field $ \Wb\in\tangentR $.}
\label{fig:schematic}
\end{figure}

Field quantities may be given \wrt\ the local parameter space $ \U $.
A vector field $ \Wb\in\tangentR $ may hence be given by
\begin{align*}
	\Wb=\begin{bmatrix}
                W^1\\ W^2 \\ W^3
            \end{bmatrix}:\quad \U \rightarrow \R^3:\quad (y^1,y^2) \mapsto \Wb(y^1,y^2)\in\tangent_{\para(y^1,y^2)}\R^3 \equiv \R^3 \formComma
\end{align*}
which trivially fulfills the restriction to $ \surf $.
Let $\eps>0$ be a small parameter and assume the local Euclidean norm of $\Wb$ to be finite,
so that $ \para+\eps\Wb $ provides a parameterization for another surface $ \surf_{\eps\Wb} $.
The surface $ \surf_{\eps\Wb} $ is hence a finite deformation of the surface $ \surf $ in the direction of $ \Wb\in\tangentR $, see figure \ref{fig:schematic}.
Even though $ \Wb $ is given on the undeformed surface $ \surf $, it can be defined independently of the parameterization $ \para $ and hence need not bear the functional argument $ \para $.

Note that we could also consider higher order deformation $ \para+\eps\Wb + \landau(\eps^2) $.
However, for the purposes of this work, linearised deformations would suffice.

\begin{remark}
 Note that our formulation is parametrization-invariant. Indeed, if we choose two parametrizations: $\tilde{\para}:\tilde{\U}\rightarrow\surf$ and $\para:\U\rightarrow\surf$ and let $\phi:\tilde{\U} \rightarrow \U$ the transition map between both parameter spaces, such that $\tilde{\para}= \para\circ\phi$. The deformation direction $\Wb = W^A \eb_A= \tilde{\Wb} = \tilde{W}^A \eb_A \in\tangentR$ is given \wrt\ both parameter spaces, \ie\ $\Wb(y^1,y^2)\in\tangent_{\para(y^1,y^2)}\R^3$ and $\tilde{\Wb}(\tilde{y}^1,\tilde{y}^2)\in\tangent_{\tilde{\para}(\tilde{y}^1,\tilde{y}^2)}\R^3$ such that $W^A = \tilde{W}^A\circ\phi$. In fact, the parametrization dependence is merely a proxy for surface dependence. 
\end{remark}
  
\subsection{Notations}

In this manuscript, lowercase Latin indices $ \{i,j,k,\ldots\} $ are used for the coordinates $ y^1 $ and $ y^2 $ defined in  the domain $ \U\subset\R^2 $ of the parameterization $ \para $;
uppercase Latin indices $ \{I,J,K,\ldots\} $ are used for the Cartesian coordinates of $ \R^3 $.
Moreover, Einstein's summation convention is adopted throughout.
Arguments in square brackets indicate total functional dependencies. 
For instance, the covariant proxy components of the metric tensor on $\surf$ are given by $ g_{ij}[\para]=\delta_{IJ}(\partial_i\paraC^I)(\partial_j\paraC^J) $, where $\delta_{IJ}$ denotes the Kronecker delta symbol and $ \partial_i := \frac{\partial}{\partial y^i} $ denotes partial differentiation with respect to $y^i$.

We write $ \qb\in\tangentS[^n] $ as a shorthand for $ \qb $ being a tangential $ n $-tensor field on $\surf$, \ie\ for $(y^1,y^2)\in\U$ we have $\qb(y^1,y^2)\in\tangentStar[^n_{\para(y^1,y^2)}]{\surf}$.
Similarly, $ \Qb\in\tangentR[^n] $ is shorthand for $ \Qb $ being a $ n $-tensor field on $\surf$\textemdash with values in $\mathbb R^{3^n}$ indeed.
Note that $ \tangentS[^n] $ is a linear subbundle of $ \tangentR[^n] $, \ie\ 
$ \tangentStar[^n_{\para(y^1,y^2)}]{\surf} \le \tangentStar[^n_{\para(y^1,y^2)}]{\R^3} $ for all $ (y^1,y^2)\in\U $,
where the equality only holds for $n=0$.
For $ n = 1 $, we omit the order index, \ie\ $ \tangent{} :=  \tangent[^1]{} $.
To illustrate local vector subspace structures, we use $\le$, or $<$.
We do not distinguish between covariant and contravariant tensor fields in index-free notations, since, for a given metric, they are equal up to isomorphism by the musical operators ($\flat$, $\sharp$).
Therefore, in writing $\tangentS[^n]$, we combine the covariant and contravariant tensor orders into a single superscript $n$. 

We use bold symbols for $ n $-tensor fields, with $ n>0 $.
For instance, we write $ \qb = q^i\partial_i\para\in\tangentS $ with contravariant proxy field $ (q^1, q^2)\in(\tangentS[^0])^2 $ in the coordinate frame $ \{\partial_i\para\}_i $. 
However, scalar fields ($ n=0 $) are not written in bold symbols.
Occasionally, we use a Cartesian frame $ \{\eb_I\} $ or the so-called thin film frame $ \{\partial_1\para, \partial_2\para, \normal[\para]\} $ restricted to the surface $ \surf $,
with normal field $ \normal[\para]\bot\tangentS $.

\subsection{Inner Products}
For inner products, we use angle brackets and mark them by their domain, 
\eg\ $ \inner{\tangentS}{\qb, \rb} := g_{ij}[\para] q^i r^j $ for 
$ \rb, \qb \in \tangentS $;
$ \inner{\tangentR}{\Qb,\Rb} := \delta_{IJ} Q^I R^J $ for $ \Qb,\Rb\in\tangentR $;
$ \inner{\tangentS[^2]}{\qb, \rb} := g_{ij}[\para]g_{kl}[\para] q^{ik} r^{jl} $ for $ \rb, \qb \in \tangentS[^2] $;
$ \inner{\tangentR[^2]}{\Qb,\Rb} := \delta_{IJ}\delta_{KL} Q^{IK} R^{JL} $ for $ \Qb,\Rb\in\tangentR[^2]$.
Since the metric tensor $\gb$ is derived from the parameterization $\para$, it therefore trivially gives an isometric embedding.
The local inner products $ \inner{\tangentS[^n]}{\cdot,\cdot} $ and $ \inner{\tangentR[^n]}{\cdot,\cdot} $ are equal 
on the subbundle $ \tangentS[^n] \le \tangentR[^n] $, \ie\ $ \inner{\tangentS}{\qb, \rb} = \inner{\tangentR}{\qb, \rb}$ for $ \rb, \qb \in \tangentS $.
The global inner product is defined as an $ \hil $-product derived from the local one
\begin{align*}
    \innerH{\mathcal{T}}{\Qb,\Rb}
        &:= \int_{\surf}  \inner{\mathcal{T}}{\Qb,\Rb} \mu[\para]
          = \int_{\U} \inner{\mathcal{T}}{\Qb,\Rb} \sqrt{\vert \gb[\para] \vert} dy^1 \!\!\wedge\! dy^2
\end{align*}
for $ \mathcal{T}=\tangentS[^n]$ (\resp\  $ =\tangentR[^n] $) and $ \Qb,\Rb\in \tangentS[^n]$ (\resp\  $ \in\tangentR[^n]$)
where $ \vert \gb[\para] \vert  $ is the determinant of the metric $ \gb[\para] $
and $ \mu[\para] $ the surface area form,
\ie\ the 2-dimensional volume form.
All norms are derived from their associated inner products, \ie\ $ \|\cdot\|_{\mathcal{T}}^2 := \inner{\mathcal{T}}{\cdot,\cdot} $
for $ \mathcal{T}\in\{ \tangentS[^n], \tangentR[^n], \hilbert{\tangentS[^n]}, \hilbert{\tangentR[^n]} \} $.

\subsection{Geometric Quantities}
The covariant proxy of the metric tensor $\gb$, also known as the first fundamental form, is given by $ g_{ij}[\para] = \inner{\tangentS}{\partial_i\para,\partial_j\para}\in\tangentS[^0] $. The normal field $ \normal[\para]\in\tangentR $ can be calculated by 
\begin{align*}
   \sqrt{\vert \gb[\para] \vert}\normal[\para] 
        &= \partial_1\para \times \partial_2\para 
         = \tensor{\epsilon}{^I_{JK}}\partial_1\paraC^J\partial_2\paraC^K\eb_{I}
\end{align*}
where $ \{\tensor{\epsilon}{^I_{JK}}\} $ is the 3-dimensional Levi-Civita symbol.
Using the 2-dimensional Levi-Civita symbols $ \{\epsilon_{ij}\} $, we define a surface Levi-Civita tensor $ \Eb[\para]\in\tangentS[^2] $ by its covariant proxy field components $ E_{ij}[\para] := \sqrt{\vert \gb[\para] \vert}\epsilon_{ij} $.\footnote{Note that even though the Levi-Civita symbol $ \{\epsilon_{ij}\} $ does not define a tensor, $ \Eb[\para]$ is a tensor indeed.}
As a bilinear map $ \tangentS\times\tangentS \rightarrow \R $, this tensor field equals the differential 2-form $ \mu[\para] $;
and as a linear map $ \tangentS[_1]\rightarrow\tangentS[_1] $ of differential 1-forms, it equals the Hodge-star operator.
There is a relation between the Levi-Civita tensor field and the $ \R^3 $-cross product: $ \qb\times\rb = (\qb\Eb[\para]\rb)\normal[\para] $ for all 
$\qb,\rb\in\tangentS$.

The covariant proxy components of the second fundamental form are given by
\begin{align}
    \shopC_{ij}[\para] 
        &:= \inner{\tangentR}{\partial_i\partial_j\para, \normal[\para] }
         = -\inner{\tangentS}{\partial_i\para, \partial_j\normal[\para] }\formPeriod
    \label{eq:shape_operator}
\end{align}
The associated 2-tensor field $ \shop[\para]\in\tangentS[^2] $ is symmetric; and as a linear map $ \tangentS\rightarrow\tangentS $, it is also known as the shape-operator. 
We use these terminologies interchangeably, since they are equal up to musical isomorphisms.
Two scalar quantities can be derived from the second fundamental form: the mean and the Gaussian curvatures $\meanc[\para],\gaussc[\para]\in\tangentS[^0]$, \resp they are defined as follows
\begin{align*}
    \meanc[\para]
        &:= \Tr\shop[\para] = \shopC^i_i[\para] \formComma
    \quad\quad\gaussc[\para]
        :=  \vert \{ \shopC^i_j \} \vert = \frac{\vert \{ \shopC_{ij} \} \vert}{\vert \gb \vert} \formComma
\end{align*}
where the determinant is given for the associated proxy matrix.

\subsection{Tangential Projection}
Since $ \tangentS[^n] $ is a subbundle of $ \tangentR[^n] $, an orthogonal projection 
$ \proj_{\tangentS[^n]}: \tangentR[^n] \rightarrow \tangentS[^n] $ may be uniquely defined as follows
\begin{align*}
    \proj_{\tangentS[^n]} \Qb 
        &= Q^{J_{1} \ldots J_{n}} 
                \left( \prod_{\alpha=1}^{n} (\delta_{J_{\alpha}}^{I_{\alpha}} - \normalC^{I_{\alpha}}[\para] \normalC_{J_{\alpha}}[\para]) \right)
                \bigotimes_{\alpha=1}^{n} \eb_{I_{\alpha}}\\
        &= \inner{\tangentR[^n]}{ \Qb, \bigotimes_{\alpha=1}^{n} \partial_{i_{\alpha}}\para }  
            \left( \prod_{\alpha=1}^{n} g^{i_{\alpha}j_{\alpha}}[\para] \right)
            \bigotimes_{\alpha=1}^{n} \partial_{j_{\alpha}}\para
\end{align*}
for all $ \Qb\in\tangentR[^n] $.
The expression in the most inner brackets of the first row defines the symmetric tangential tensor field 
$ \projb_{\surf} := (\delta^{IJ} - \normalC^I[\para]\normalC^J[\para]  )\eb_{I}\otimes\eb_{J}\in\tangentS[^2] $.
Note that  $ \proj_{\tangentS[^2]}\Qb = \projb_{\surf}\Qb\projb_{\surf}$ for all $ \Qb\in\tangentS[^2] $.
The restriction to $\tangentS$ of the tensor field $ \projb_{\surf} $ also describes the tangential field identity map $ \Id_{\tangentS}:\tangentS\rightarrow\tangentS $, 
since $ [\projb_{\surf}\vert_{\tangentS}]^i_j = \delta^i_j $ in local coordinates.
Therefore, we write also $ \Id_{\tangentS}=\projb_{\surf} $ synonymously in appropriate situations.

The space of symmetric and trace-free tangential 2-tensors $ \Qcal^2\surf $ is a subbundle of $ \tangentS[^2] $,
and the associated orthogonal projection $ \proj_{\Qcal^2\surf}:  \tangentS[^2] \rightarrow \Qcal^2\surf $  is given by
$
    \proj_{\Qcal^2\surf}\qb 
        := \frac{1}{2}\left( \qb + \qb^T - (\Tr\qb)\Id_{\tangentS} \right)
$
and is surjective.

\subsection{Spatial Derivatives}
We use the covariant derivative $ \nabla:\tangentS[^n]\rightarrow\tangentS[^{n+1}] $ defined by
\begin{align}
    [\nabla\qb]_{i_1\ldots i_n k} := q_{i_1\ldots i_n \vert k}
        &:= \partial_k q_{i_1\ldots i_n}
                - \sum_{\alpha=1}^{n} \Gamma_{k i_{\alpha}}^j[\para] q_{i_1 \ldots i_{\alpha-1} j i_{\alpha+1}  \ldots i_n}
    \label{eq:covariant_derivative}
\end{align}
on fully covariant proxy components, where $ \Gamma_{ki}^j[\para] = g^{jl}[\para]\Gamma_{kil}[\para]$ are the Christoffel symbols of second kind
and $ \Gamma_{ijk}[\para] $ of first kind given by
\begin{align}
    \Gamma_{ijk}[\para]
        &= \frac{1}{2} \left( \partial_i g_{jk}[\para] + \partial_j g_{ik}[\para] - \partial_k g_{ij}[\para] \right)
         = \inner{\tangentR}{\partial_i\partial_j\para, \partial_k\para } \formPeriod
    \label{eq:Christoffel_first_kind}
\end{align}
Note that for raising an index in the left-hand-side in \eqref{eq:covariant_derivative}, the sign and index order of the corresponding Christoffel symbol changes in the right-hand-side by virtue of metric compatibility, \ie\ $ \partial_k g_{ij}[\para] = \Gamma_{kij}[\para] + \Gamma_{kji}[\para] $.
For instance, the covariant derivative of a tangential vector field $ \qb\in\tangentS $ reads
\begin{align*}
    \tensor{q}{^i_{\vert k}}
        &:=g^{ij}[\para] q_{j \vert k}
         = g^{ij}[\para] \left( \partial_k ( g_{jl}[\para] q^l ) - \Gamma_{kj}^l[\para] q_l \right) \\
        &= \delta^i_l\partial_k q^l + g^{ij}[\para]q^l\left( \Gamma_{kjl}[\para] + \Gamma_{klj}[\para] \right)  - \Gamma_{kj}^l[\para] q_l
         = \partial_k q^i + \Gamma_{kj}^i[\para] q^j \formPeriod
\end{align*}

We identify the usual covariant $ \R^3 $-derivative restricted to the surface by 
\begin{align*}
    \nabla_{\R^3}:= \delta^{IK}\partial_I(\cdot)\vert_{\surf} \otimes \eb_K: \tangentR[^n] \rightarrow \tangentR[^{n+1}] \formPeriod
\end{align*}
Due to the lack of information\textemdash and arbitrariness\textemdash along the normal direction\footnote{Since in $\tangentR[^n]$, the tensor fields are only defined on $\surf$.}, this derivative is undetermined.
But we can define the unique surface and tangential derivatives, \resp, from it by
\begin{align}
    \nabla_{\surf} &:= \nabla_{\R^3}\cdot\projb_{\surf}: \tangentR[^n] \rightarrow \tangentR[^n]\otimes\tangentS \formComma
    \label{eq:surface_derivative}\\
    \nabla_{\tangentS[^{n+1}]} &:= \proj_{\tangentS[^{n+1}]}\circ\nabla_{\R^3}: \tangentR[^n] \rightarrow \tangentS[^{n+1}] \formPeriod
    \label{eq:tangential_derivative}
\end{align}
Note that $ \nabla_{\tangentS[^{n+1}]}\vert_{\tangentS[^n]} = \nabla $
while $\forall\qb\in\tangentS$: $\nabla_{\mathcal{S}}\qb=\nabla_{\tangentS[^2]}\qb + \normal\otimes\shop\qb$.

The curl-operator $ \rot:\tangentS \rightarrow \tangentS[^0] $ on tangential vector fields $ \qb\in\tangentS $ is given by
\begin{align*}
     \rot\qb := -\inner{\tangentS[^2]}{\nabla\qb, \Eb[\para]},
\end{align*}
and the relation to the antisymmetric part of the covariant derivative
is $ \frac{1}{2}(\nabla\qb - (\nabla\qb)^T) = -\frac{\rot\qb}{2}\Eb[\para] $.

The covariant divergence $ \div: \tangentS[^n] \rightarrow \tangentS[^{n-1}] $ is defined as the $ \hil $-adjoint $ \nabla^\adj $ of $ \nabla $ and it is given in components by contracting the last two indices of its image
\begin{align*}
    \div\qb 
        &:= -\nabla^{\adj}\qb
          = \Tr_{(n,n+1)}\nabla\qb
          = \tensor{q}{^{i_1 \ldots i_{n-1} k}_{\vert k}} \bigotimes_{\alpha=1}^{n-1} \partial_{i_{\alpha}}\para\,,
\end{align*}
for all $ \qb\in\tangentS[^n] $.
The  $\hil$-adjoint $ \nabla_{\tangentS[^2]}^\adj $ of $ \nabla_{\tangentS[^2]} $ defines the tangential divergence
\begin{align}
    \div_{\tangentS[^2]}
        &:= -\nabla_{\tangentS[^2]}^\adj
          = \div(\cdot) + \inner{\tangentS[^2]}{\shop[\para],\cdot}\normal[\para]: 
            \quad\tangentS[^2] \rightarrow \tangentR \,.
     \label{eq:tangential_divergence}
\end{align}
For all $ \sigmab\in\tangentS[^2] $ and $ \etab\in\tangentS $, the  $\hil$-adjoint $ \nabla_{\surf}^\adj $ of $ \nabla_{\surf} $ defines the surface divergence
\begin{align}
    \div_{\surf}(\sigmab + \normal[\para]\otimes\etab)
        &:=  -\nabla_{\surf}^\adj (\sigmab + \normal[\para]\otimes\etab) 
           =  (\Tr_{(2,3)}\circ\nabla_{\surf})(\sigmab + \normal[\para]\otimes\etab) \notag\\
         &= \div\sigmab - \shop[\para]\etab + (\inner{\tangentS[^2]}{\shop[\para],\sigmab} + \div\etab)\normal[\para]\,,
    \label{eq:surface_divergence}
\end{align}
where integration by parts in \eqref{eq:tangential_divergence} is done \wrt\ $ \nabla $ and not $ \nabla_{\R^3} $, since the global inner product is given by
$ \hilbert{\tangentS[^2]} $, \resp\ $ \hilbert{\tangentR\otimes\tangentS} $.
Hence, it follows that $ \div_{\surf}\vert_{\tangentS[^2]} = \div_{\tangentS[^2]} $.
Note that $ \div_{\tangentS[^2]} $ cannot be achieved through applying the trace on its related gradient, 
contrarily to the divergences $ \div_{\surf} $ and $ \div $.

\subsection{Gradient of Vector Fields}\label{sec:gradient_of_W}

We now consider the gradient of vector fields $ \Wb\in\tangentR $ and distinguish its tangential, symmetric and antisymmetric parts.
In most cases, $ \Wb $ is used as the deformation direction field.
Since $ \Wb $ is an $ \R^3 $-quantity, though restricted to $ \surf $, we start with the $ \R^3 $-gradient $ \nabla_{\R^3}\Wb\in\tangentR[^2] $
\wrt\ the embedding space.  
This includes the derivative in normal direction, which is clearly not determined due to the restriction $ \vert_{\surf} $, but we do not need it.

The pure tangential part is $ \Gb[\para,\Wb]:= \nabla_{\tangentS[^2]}\Wb \in\tangentS[^2] $, according to definition \eqref{eq:tangential_derivative},
and its covariant proxy components yield
\begin{align}
    G_{ij}[\para,\Wb] 
        &= \inner[1]{\tangentS[^2]}{\nabla_{\tangentS[^{2}]}\Wb , \partial_i\para \otimes \partial_j\para}
         = \inner[1]{\tangentR}{\partial_j\Wb, \partial_i\para} \formPeriod
    \label{eq:tangential_part_of_gradW}
\end{align}
An orthogonal decomposition $ \Wb=\wb[\para] + \wnor[\para]\normal[\para] $ with tangential part $ \wb[\para]\in\tangentS $ 
and scalar normal part $ \wnor[\para]\in\tangentS[^0] $ results in
\begin{align}
    \partial_j\Wb
        &= (\partial_j w^k[\para])\partial_k\para 
            + w^k[\para]\partial_j\partial_k\para
            + \wnor[\para]\partial_j\normal[\para] 
            + (\partial_j\wnor[\para])\normal[\para]\formPeriod
     \label{eq:tangential_derivative_of_W_components}
\end{align}
Therefore, we obtain by means of \eqref{eq:shape_operator}, \eqref{eq:covariant_derivative} and \eqref{eq:Christoffel_first_kind}, that
\begin{align*}
    G_{ij}[\para,\Wb]
        &= g_{ik}[\para]\partial_j w^k[\para] 
          + w^k[\para]\Gamma_{jki}[\para]
          - \wnor[\para]\shopC_{ij}[\para]\\
        &= w_{i\vert j}[\para] - \wnor[\para]\shopC_{ij}[\para]\formPeriod
\end{align*}
The covariant proxy components of the symmetric part $ \Sb[\para,\Wb]:=\frac{1}{2}(\Gb[\para] + \Gb^T[\para])\in\tangentS[^2] $
are given by
\begin{align*}
    S_{ij}[\para,\Wb]
        &= \frac{1}{2}\left(w_{i\vert j}[\para] + w_{j\vert i}[\para]\right) -  \wnor[\para]\shopC_{ij}[\para]\,,
\end{align*}
as a consequence.
In contrast, the antisymmetric part $ \Ab[\para,\Wb]:= \frac{1}{2}(\Gb[\para] - \Gb^T[\para])\in\tangentS[^2] $ becomes
\begin{align*}
    A_{ij}[\para,\Wb]
            &= \frac{1}{2}\left(w_{i\vert j}[\para] - w_{j\vert i}[\para]\right)
             = \frac{\rot\wb[\para]}{2} E_{ji}[\para]\,,
\end{align*}
and does not depend on normal components given by $ \wnor[\para] $.
In table \ref{tab:gradients_of_W} we give a summary of these three tangential 2-tensor fields in index-free notations. 
\begin{table}
\centering
\renewcommand{\arraystretch}{1.5}
\begin{tabular}{|c|c|c|}
    \hline 
        & covariant surface calculus 
            & $\R^3$-calculus \\
    \hline
    $\!\Gb[\para,\Wb]\!$ 
        & $\nabla\wb[\para] - \wnor[\para]\shop[\para] $ 
            & $\proj_{\tangentS[^2]}(\nabla_{\R^3}\Wb)$\\
    \hline
    $\!\Sb[\para,\Wb]\!$ 
        & $\frac{1}{2}\left(\nabla\wb[\para] + (\nabla\wb[\para])^T \right)- \wnor[\para]\shop[\para] $ 
            & $\!\frac{1}{2}\proj_{\tangentS[^2]}(\nabla_{\R^3}\Wb + (\nabla_{\R^3}\Wb)^T)\!$\\
    \hline
    $\!\Ab[\para,\Wb]\!$
        & $\!\frac{1}{2}\left(\nabla\wb[\para] - (\nabla\wb[\para])^T \right) = -\frac{\rot\wb[\para]}{2}\Eb[\para]\!$
            & $\!\frac{1}{2}\proj_{\tangentS[^2]}(\nabla_{\R^3}\Wb - (\nabla_{\R^3}\Wb)^T)\!$\\
    \hline
\end{tabular}
\caption{Fully tangential part of $ \nabla_{\R^3}\Wb $ and its orthogonal decomposition $ \Gb[\para,\Wb]= \Sb[\para,\Wb] + \Ab[\para,\Wb]\in\tangentS[^2] $
            into symmetric part $ \Sb[\para,\Wb]\in\tangentS[^2] $ and antisymmetric part $ \Ab[\para,\Wb]\in\tangentS[^2] $
            in terms of $ \R^3 $-calculus \wrt\ the embedding space (right) and a covariant surface calculus (middle),
            where $ \Wb=\wb[\para] + \wnor[\para]\normal[\para]\in\tangentR $ is orthogonally decomposed into tangential field $ \wb[\para]\in\tangentS $ and
            scalar normal field $ \wnor[\para]\in\tangentS[^0] $.
            }
\label{tab:gradients_of_W}
\end{table}

From \eqref{eq:tangential_derivative_of_W_components}, it follows that the normal component of the derivative in tangential direction is given by
\begin{align}
    \inner{\tangentR[^2]}{ \nabla_{\R^3}\Wb, \normal[\para]\otimes\partial_i\para}
        &= \inner{\tangentR}{ \partial_i\Wb , \normal[\para]}\notag\\
        &=  (\partial_i\wnor[\para])\left\| \normal[\para] \right\|^2_{\tangentR}
            +w^k[\para] \inner{\tangentR}{ \partial_i\partial_k\para , \normal[\para]}\notag\\
        &= \partial_i \wnor[\para] + \shopC_{ik}[\para] w^k[\para]\formPeriod
    \label{eq:normal_gradient_of_W}
\end{align}
Using \eqref{eq:tangential_part_of_gradW}, the surface derivative \eqref{eq:surface_derivative} of $ \Wb $
becomes
\begin{align}
    \nabla_{\surf}\Wb 
        &:= \Gb[\para,\Wb] + \normal[\para]\otimes\bb[\para,\Wb] \in \tangentR\otimes\tangentS \formComma 
    \label{eq:surface_derivative_of_W} \\
    \text{where}\quad\bb[\para,\Wb] &:= \nabla\wnor[\para] + \shop[\para]\wb[\para] \in\tangentS \formPeriod 
    \label{eq:b_definition}
\end{align}

We also need the $\hil$-adjoint 
$ \Psib^{\dadj}[\para]:\tangentS[^2] \rightarrow \tangentR $ of $ \Psib\in\{\Gb,\Gb^T,\Sb,\Ab\} $, which is implicitly defined by
\begin{align*}
    \forall \rb\in\tangentS[^2]:\quad 
    \innerH{\tangentR}{\Psib^{\dadj}[\para](\rb), \Wb}
        := \innerH{\tangentS[^2]}{\rb, \Psib[\para,\Wb]} \formPeriod 
\end{align*}
Since $ \Gb[\para,\Wb] = \nabla_{\tangentS[^2]}\Wb $,
tangential divergence \eqref{eq:tangential_divergence} yields
\begin{align}
    \Gb^\dadj[\para](\rb)
        &= -\div_{\tangentS[^2]}\rb
        &&= -\left(\div\rb + \inner{\tangentS[^2]}{\rb,\shop[\para]}\normal[\para]\right)\notag\\
    (\Gb^T)^\dadj[\para](\rb)
        &= -\div_{\tangentS[^2]}\rb^T
        &&= -\left(\div\rb^T + \inner{\tangentS[^2]}{\rb,\shop[\para]}\normal[\para]\right)\notag\\
    \Sb^\dadj[\para](\rb)
        &= -\div_{\tangentS[^2]}\frac{\rb + \rb^T}{2}
        &&= -\left(\div\frac{\rb + \rb^T}{2} + \inner{\tangentS[^2]}{\rb,\shop[\para]}\normal[\para]\right)\notag\\
    \Ab^\dadj[\para](\rb)
        &= -\div_{\tangentS[^2]}\frac{\rb - \rb^T}{2}
        &&= -\div\frac{\rb - \rb^T}{2}
    \label{eq:deformation_adjoints}
\end{align}


\section{Gauges of Surface Independence, $ \hil $-Gradient and Flows}\label{sec:goi_main}

To calculate a consistent $ \hil $-gradient flow, it is mandatory to clarify the role of independence between the degrees of freedom of the underlying system. We quantify the independence of the surface for tangential tensor fields by various gauges of surface independence. In order to do that, we introduce a variety of deformation derivatives, which can be seen as local variations of tensor fields \wrt\ the surface. A gauge of surface independence is fulfilled if and only if the associated deformation derivative vanishes.

We treat scalar fields in subsection \ref{scalar_fields}. Even if the proceeding here seems a bit superfluous, it gives us the opportunity to introduce basic concepts and to create an awareness of emerging problems in the course of surface variations. In subsection \ref{sec:vector_fields}, we take a closer look at this for tangential vector fields. Afterwards, we give a short summary of the results for tangential 2-tensor fields in subsection \ref{sec:tangential_tensor_fields}. Generalizations to tangential n-tensor fields can be done along the same lines.

In each subsection, we discuss consequences for energy variations, the resulting $ \hil $-gradients, and their flows. Differentiating energies that depend on tensor fields by the underlying spatial space could bear some uncertainties. Physical systems are often only defined on a currently given space, possibly also on a moving space if the system is time-dependent. However, in many energy techniques, more information about the behavior of energy variation is essential. To calculate global gradients or total differentials of energies, we have to vary the energy in space arbitrarily and instantaneously, though we do not know how the energy acts in a small vicinity of the underlying space, since this is not a part of the prescribed physical system. As we see in each subsection, all uncertainties that arise from arbitrary spatial variations can be determined by the gauges of surface independence.

\subsection{Scalar Fields}\label{scalar_fields}
Let us consdier a smooth scalar field $f$ on the 2-dimensional smooth surface $\mathcal S$ parameterized by $\para$:
\begin{align}\label{eq:scalar_fields}
    f[\para]:\quad \U \rightarrow \R;\quad (y^1,y^2) \mapsto f[\para](y^1,y^2)\in\tangent^0_{\para(y^1,y^2)}\surf  \formPeriod 
\end{align}  
Recall that arguments in squared brackets indicate a total functional dependency. 
This means that $ f $ could also depend partially on derivatives of $ \para $.
For instantce, covariant proxy components of the metric tensor yield 
$ g_{ij}[\para] =  \inner{\tangentR}{\partial_i\para , \partial_j\para} \in\tangentS[^0] $.
It should be noted that the parameterization $\para$ remains only an arbitrary proxy for the surface $\surf$.
If we consider an alternative parameterization $\tilde{\para}:\tilde{\U} \rightarrow \surf$, accompanied by a  transition map $\phi:\tilde{\U} \rightarrow \U$ such that  $\tilde{\para}=\para\circ\phi$, we can define an equivalent scalar field $\tilde{f}[\tilde{\para}]\in\tangentS[^0]$ as follows: $\tilde{f}[\tilde{\para}](\tilde{y}^1, \tilde{y}^2) := (f[\tilde{\para}\circ\phi^{-1}]\circ\phi)(\tilde{y}^1, \tilde{y}^2)$.
This can also be generalized when the images of both parameterizations overlap to an open subset of the surface.

\subsubsection{Gauge of Surface Independence}
For any vector field $\Wb\in\tangentR$ and a small parameter $\eps > 0$, $ \para+\eps\Wb $ provides a parameterization  in $\mathbb R^3$ for a $\eps$-perturbed $ \surf $ in the $\Wb$ direction; and we may define the \emph{deformation derivative} $ \gauge{\Wb} f [\para] $ in the direction of $ \Wb\in\tangentR $:
\begin{align}
    \gauge{\Wb} f [\para](y^1,y^2)
        :=\lim_{\eps\rightarrow 0 } \frac{f[\para + \eps\Wb](y^1,y^2) - f[\para](y^1,y^2)}{\eps}\in\tangent^0_{\para(y^1,y^2)}\surf\formComma
\label{eq:deformation_derivativ_scalar}
\end{align}
in such a way that it is well-defined at all $ (y^1,y^2)\in\U $.
As mentioned above, $f[\para]$ is invariant \wrt\ the choice of parameterization. The same applies to the deformation derivative.
For an alternative parameterization $\tilde{\para}$ and an equivalent scalar field $\tilde{f}[\tilde{\para}]\in\tangentS[^0]$, such that $\tilde{f}[\tilde{\para}](\tilde{y}^1, \tilde{y}^2)= f[\para](y^1,y^2)$ at every location $\tilde{\para}(\tilde{y}^1, \tilde{y}^2) = \para(y^1,y^2)$,
is also $\tilde{f}[\tilde{\para}+\eps\tilde{\Wb}](\tilde{y}^1, \tilde{y}^2)= f[\para + \eps\Wb](y^1,y^2)$ valid for an equivalent deformation directions $\tilde{\Wb}(\tilde{y}^1, \tilde{y}^2) = \Wb(y^1,y^2) \in \tangent_{\tilde{\para}(\tilde{y}^1, \tilde{y}^2)}\R^3 = \tangent_{\para(y^1,y^2)}\R^3$.
Therefore, $ \gauge{\tilde{\Wb}} \tilde{f}[\tilde{\para}](\tilde{y}^1, \tilde{y}^2) = \gauge{\Wb} f [\para](y^1,y^2) $ holds locally and 
$ \gauge{\tilde{\Wb}} \tilde{f}[\tilde{\para}] = ( \gauge{\tilde{\Wb}\circ\phi^{-1}} f [\tilde{\para}\circ\phi^{-1}])\circ\phi$ \wrt\ the transition map $\phi:\tilde{\U} \rightarrow \U$ introduced above.
The deformation derivative \eqref{eq:deformation_derivativ_scalar} measures the linear change of a scalar field 
\wrt\ a change of the underlying surface.

\begin{definition}[Gauge of Surface Independence on Scalar Fields]\label{def:goi_scalar}
    A scalar field $ f[\para]\in \tangentS[^0] $ fulfills the gauge of surface independence $ \gauge{}f=0 $, if and only if
    $ \gauge{\Wb} f [\para] = 0 $ is valid everywhere in $ \U $ for all $ \Wb\in\tangentR $.
\end{definition}
Using a Taylor expansion at $ \eps=0 $ yields
\begin{align}
    f[\para + \eps\Wb](y^1,y^2) &= f[\para](y^1,y^2) + \eps\gauge{\Wb} f [\para](y^1,y^2) + \landau(\eps^{2})\formComma
    \label{eq:taylor_f}
\end{align}
which shows that the gauge $ \gauge{}f=0 $ is less restrictive for determining $f[\para + \eps\Wb]$ sufficiently than the example of pushforward above.
Note that $ \gauge{}f=0 $ does not imply $f[\para]\equiv 0$, since every scalar field $f[\para]:(y^1,y^2)\mapsto \tilde{f}(y^1,y^2)\neq 0$ 
would bear this gauge of surface independence.

In the following, we omit the partial coordinates dependency $ (y^1,y^2) $ as much as possible in the notion for field quantities.
Unless stated otherwise, field quantities and operators are always defined pointwise for all $ (y^1,y^2) \in\U$. 

The gauge of surface independence is important as part of assumptions which should be forced a priori if the steepest decent of an energy is of interest \wrt\ scalar fields and the surface.
Every scalar field obeying the gauge $ \gauge{}f=0 $ can be seen as independent of surface deformations, thus it is a natural assumption for several variation principles,
where the underlying surface is a degree of freedom, rather than a static background for the physical model.
For instance, the proxy field of the metric tensor generally fails this gauge componentwise, since for $ f[\para]=g_{ij}[\para] $, it holds
\begin{align}
    \gauge{\Wb}g_{ij}[\para] 
        &= \inner{\tangentR}{\partial_i \Wb , \partial_j\para} + \inner{\tangentR}{\partial_j \Wb , \partial_i\para}\formPeriod
\label{eq:deformation_g_I}
\end{align}
$\gauge{\Wb} f [\para] = \gauge{\Wb}g_{ij}[\para] = 0$ is valid only if $\Wb$ is the linear direction of a rigid body deformation of the surface.
Therefore, minimizing a pure geometric energy $ \energy=\energy[\surf[\para], \{g_{ij}[\para]\}] $ \wrt\ $ \surf[\para] $ and all $ g_{ij}[\para] $ simultaneously is not recommended
and would possibly give an inappropriate $ \hil $-gradient, which may for instance lead to an overdetermined gradient flow.
Moreover, the surface as well as  the metric tensor are even fully determined by the parameterization $ \para $ and therefore fully dependent variables,
hence we could formulate the same energy just as $ \energy=\energy[\para] $.

\subsubsection{Energy Variations}

For now, we do not assume specific dependencies on $\para$ for scalar fields and discuss the role of the gauge of surface independence at the end of this section.
Considering an energy $ \energy[\para,f[\para]] $ depending on the surface by $\para$ and a scalar field $ f[\para]\in\tangentS[^0] $, the total and partial 
variation\footnote{In this work, we distinguish “partial” and “total” derivatives: Total derivatives affect the semantic change of the argument, whereas partial derivatives are depending on the syntax of their arguments.} 
of the energy \wrt\ $ \para $, evaluated at $\para$, are respectively given by
\begin{align*}
     \inner[4]{\hilbertR}{\frac{\del\energy}{\del\para} , \Wb} 
            &:= \lim_{\eps\rightarrow 0 } \frac{\energy[\para + \eps\Wb, f[\para + \eps\Wb]] - \energy[\para, f[\para]]}{\eps} \in\R \formComma\\
     \inner[4]{\hilbertR}{\frac{\partial\energy}{\partial\para} , \Wb} 
                 &:= \lim_{\eps\rightarrow 0 } \frac{\energy[\para + \eps\Wb, f[\para]] - \energy[\para, f[\para]]}{\eps} \in\R
\end{align*}
and the variation \wrt\ $ f $ by
\begin{align*}
     \inner[4]{\hilbertS[^0]}{\frac{\del\energy}{\del f},  r} \hspace{-0pt}
            &:= \lim_{\eps\rightarrow 0 } \frac{\energy[\para, f[\para] + \eps r] - \energy[\para, f[\para]]}{\eps} \in\R
\end{align*}
for all $ \Wb\in\tangentR $ and $ r\in\tangentS[^0] $.
Taylor expansion \eqref{eq:taylor_f} and an additional expansion of $ \energy $ yields
\begin{align*}
    \energy[\para + \eps\Wb, f[\para + \eps\Wb]]
        &= \energy[\para + \eps\Wb, f[\para]] + \eps \inner[4]{\hilbertS[^0]}{\frac{\del\energy}{\del f},  \gauge{\Wb}f[\para]} 
        \hspace{-10pt}+ \landau(\eps^2)\formPeriod
\end{align*}
As a consequence we can relate the different variations by
\begin{align*}
    \inner[4]{\hilbertR}{\frac{\del\energy}{\del\para} , \Wb}
        &= \inner[4]{\hilbertR}{\frac{\partial\energy}{\partial\para} , \Wb} 
          + \inner[4]{\hilbertS[^0]}{\frac{\del\energy}{\del f},  \gauge{\Wb}f[\para]} \formPeriod
\end{align*}
Therefore, assuming the gauge $ \gauge{}f=0 $, the partial and total variation of $\energy$ \wrt\ $ \para $ would be equal, as one would expect for energies with independent variables.

\begin{remark}
 Note that the attribute ``well-defined'' is part of the definition of the deformation derivative \eqref{eq:deformation_derivativ_scalar} and not a conclusion, \ie\ $ f[\para + \eps\Wb] $ has to be known sufficiently
so that $\gauge{\Wb} f [\para]$ and $\frac{\del\energy}{\del\para}$ can be seen as adequately defined G\^ateaux derivatives.
A usual way to archive this is to push $f[\para]$ forward \wrt\ the map $ \surf \rightarrow \surf_{\eps\Wb} $, 
\eg\ by stipulating
$ f[\para + \eps\Wb](y^1,y^2) := f[\para](y^1,y^2) $.
This way to determine  $ f[\para + \eps\Wb] $ can be seen as a sufficient assumption to state that the scalar field is independent of surface deformations.
However, we can also go the other way around and stipulate that the deformation derivative has to vanish.
\end{remark}

\subsubsection{$ \hil $-Gradient Flow}
To formulate a $ \hil $-gradient flow, we need independence of the degrees of freedoms $ \para $ and $ f[\para] $,
\ie\ we stipulate the gauge of surface independence $ \gauge{}f=0 $.
In order to describe an observer-invariant rate for a scalar field $ f[\para] $ we use the material time derivative, given in \cite{Nitschke_2022} for instance.
We write $ \dot{f}[\para] $ for this rate.
Observer-invariance means that the time-depending observer parameterization does not have to be the same as the parameterization describing the material as long as both represent the same moving surface. For more details, see \cite{Nitschke_2022}.  
Eventually, the strong $ \hil $-gradient flow of an energy $\energy = \energy[\para,f[\para]] $ reads
\begin{align}
    \frac{d}{dt}\para &= -\lambda\frac{\del\energy}{\del\para} = -\lambda\frac{\partial\energy}{\partial\para}\formComma
    \quad\quad\dot{f}[\para] = -\lambda\frac{\del\energy}{\del f} \formComma
\end{align}
with mobility coefficient $ \lambda>0 $. 
The energy is dissipative under this gradient flow, since
\begin{align*}
    \frac{d}{dt}\energy 
        &= -\lambda \left( \left\| \frac{\del\energy}{\del\para} \right\|^2_{\hilbertR} + \left\| \frac{\del\energy}{\del f} \right\|^2_{\hilbertS[^0]} \right) \le 0 \formPeriod
\end{align*}
Many examples exist where this is used intuitively without explicitly considering the gauge of surface independence $ \gauge{}f=0 $, see section \ref{sec:example_scalar}.

\subsection{Tangential Vector Fields}\label{sec:vector_fields}

In this section we investigate the impact of surface deformations on tangential vector fields $\qb[\para]\in\tangentS $.
Similarly to scalar fields \eqref{eq:scalar_fields}, they are defined pointwise by
\begin{align*}
    \qb[\para]:\quad (y^1,y^2) \mapsto \qb[\para](y^1,y^2)\in\tangent_{\para(y^1,y^2)}\surf <  \tangent_{\para(y^1,y^2)}\R^3 \formPeriod
\end{align*}
Due to syntactically different, but isomorphic, representation possibilities for tangential vector fields, there are different natural ways to define the degrees of freedom.
Eventually, these differences subsequently result in varying gauges of surface independence.
Two representations for tangential vector fields $ \qb[\para]\in\tangentS $ are well-known in differential geometry and tensor analysis.
One is the contravariant proxy vector field $ \{q^i[\para]\}\in(\tangentS[^0])^2 $, which assembles the vector field $\qb[\para]$ \wrt\ a given frame.
We employ $\qb[\para]=q^i[\para]\partial_i\para$ in this paper.
The other one is the covariant proxy vector field $ \{q_i[\para]\}\in(\tangentS[^0])^2 $, which generates the linear map $\inner{\tangentS}{\qb[\para], \cdot}:\tangentS\rightarrow\tangentS[^0]$ dual to $\qb[\para]$, where the dual frame could be derived from the primal one.
This means in our case that $ \inner{\tangentS}{\qb[\para], \cdot} = q_i[\para] \inner{\tangentS}{\partial^i\para, \cdot} $ with $\partial^i\para := g^{ij}[\para] \partial_j\para$.
Both are isomorphically related by $ q_i[\para] = g_{ij}[\para] q^j[\para] $, \resp\ by $ q^i[\para] = g^{ij}[\para]q_j[\para] $.
In this context, it is worthwhile to recall $g_{ij}[\para]=\inner{\tangentR}{\partial_i\para, \partial_j\para} $
and its proxy matrix field inverse given by $ g^{ik}[\para]g_{kj}[\para]= \delta^i_j  $ implicitly.
Since proxy vector fields are only the pointwise Cartesian product of scalar fields, we could use the deformation derivative \eqref{eq:deformation_derivativ_scalar} componentwise
and the gauge of surface independence \wrt\ Definition \ref{def:goi_scalar} holds also componentwise. 
But the resulting gauges $ \gauge{}q^i=0 $ and $ \gauge{}q_i=0 $ are obviously not the same, since in general the metric tensor depends on the surface as we have seen in \eqref{eq:deformation_g_I}.
In this section we add two more gauges, name them, and relate them all to each other.
Afterwards, we discuss consequences for energy variations and $ \hil $-gradient flows \wrt\ these gauges.

\subsubsection{Gauges of Surface Independence}
By recognizing the embedding space, tangential vector fields $ \qb[\para]\in\tangentS < \tangentR $ lay in a linear sub-vector bundle of 
$ \tangentR = \tangentS\oplus(\tangentS[^0])\normal[\para]$.
Therefore, every tangential vector field $ \qb[\para]\in\tangentS$ can be represented in $ \tangentR $,
\ie\ it holds $ \qb[\para]= q^I[\para]\eb_I = q^i[\para]\partial_i\para $,
where $ \eb_I $ are the Cartesian base vector fields.
The advantage of using a Cartesian frame is that it is constant and defined in the whole $\tangent{\R^3}$.
Hence, also on the deformed surface $ \surf_{\eps\Wb} $ the tangential vector field $ \qb[\para + \eps\Wb]= q^I[\para + \eps\Wb]\eb_I\in\tangent{\surf_{\eps\Wb}} $ is represented with the same base vector fields 
$ \{\eb_I\} $.
Since a derivative has to obey the product rule it is justified to define 
\begin{align*}
    \gauge{\Wb} \qb[\para] 
        = \gauge{\Wb} \left( q^I[\para]\eb_I \right)
        := \left(\gauge{\Wb} q^I[\para]\right) \eb_I 
\end{align*}
on $\tangentS$ by the deformation derivative \eqref{eq:deformation_derivativ_scalar} for scalar fields.
Changing the frame to $ \{\partial_i\para\} $ yields
\begin{align}
    \gauge{\Wb}\qb[\para]
        &= \lim_{\eps\rightarrow 0 } \frac{q^I[\para+\eps\Wb] - q^I[\para]}{\eps} \eb_I\notag\\
         &= \lim_{\eps\rightarrow 0 } \frac{q^i[\para+\eps\Wb]\left( \partial_i\para + \eps\partial_i\Wb \right) - q^i[\para]\partial_i\para}{\eps} \notag\\
         &= \left( 	\gauge{\Wb}q^i[\para] \right)\partial_i\para +  q^i[\para]\partial_i\Wb\\
         &= \left( 	\gauge{\Wb}q^i[\para] \right)\partial_i\para +  (\nabla_{\surf}\Wb)\qb[\para] \in\tangentR \formPeriod
                \label{eq:deformation_derivativ_full_vector}
\end{align}
The first summand implies that this deformation derivative is determined by prescribing $ \gauge{\Wb}q^i[\para] $, 
while the second summand is already determined at the surface by $\para$, $\Wb$ and $\qb[\para]$.
By orthogonal decomposition $ \Wb=\wb[\para] + \wnor[\para]\normal[\para] $ and surface derivative \eqref{eq:surface_derivative_of_W} of $ \Wb $, 
the normal part of $ \gauge{\Wb}\qb[\para] $ is
\begin{align*}
    \inner{\tangentR}{\gauge{\Wb}\qb[\para] , \normal[\para] }
        &= \nabla_{\qb[\para]}\wnor[\para] + \qb[\para]\shop[\para]\wb[\para] \formPeriod
\end{align*}
Since this normal part, coming only from the second summand of \eqref{eq:deformation_derivativ_full_vector},  is fully determined by the surface,
it is justified to use only the tangential part of the deformation derivative given by $\gauge{\Wb}\qb[\para]  $.
With surface derivative \eqref{eq:surface_derivative_of_W} of $ \Wb $,
the covariant proxy field of the tangential part of \eqref{eq:deformation_derivativ_full_vector} yields
\begin{align*}
    \left[ \gauge{\Wb}\qb[\para] \right]_i
        &= \inner{\tangentR}{\gauge{\Wb}\qb[\para], \partial_i\para }
         = g_{ij}[\para] \gauge{\Wb}q^j[\para] + G_{ij}[\para,\Wb]q^j[\para] \formComma
\end{align*}
where $ G_{ij}[\para,\Wb] = w_{i\vert j}[\para] - \wnor[\para]\shopC_{ij}[\para] $ is the tangential derivative of $ \Wb $, see section \ref{sec:gradient_of_W}.
With this we define the \emph{material deformation derivative} $ \gaugeS[\Wb]: \tangentS \rightarrow \tangentS $ of tangential vector fields
in direction of $ \Wb\in\tangentR $ by
\begin{align}
    \gaugeS[\Wb]\qb[\para] 
        &:=  (\proj_{\tangentS}[\para] \circ \gauge{\Wb})\qb[\para]
          = g^{jk}[\para]\left[ \gauge{\Wb}\qb[\para] \right]_j \partial_k \para \notag\\
         &= ( \gauge{\Wb}q^k[\para] ) \partial_k \para +\Gb[\para,\Wb]\qb[\para]\formPeriod
         \label{eq:deformation_derivativ_vector}
\end{align}
Given a tangential vector field $\qb[\para]$ on $\surf$, its evolution on the $\eps$-perturbed surface $\surf_{\eps\Wb}$ is a priori unknown. One way of getting rid of this indeterminacy is defining the following surface gauge:
\begin{definition}[Material Gauge of Surface Independence on Tangential Vector Fields]\label{def:goi_vector}
    A tangential vector field $ \qb[\para]\in \tangentS $ fulfills the gauge of material surface independence $ \gaugeS\qb=0 $, if and only if
    $ \gaugeS[\Wb] \qb[\para] = 0 $ for all $ \Wb\in\tangentR $.
\end{definition}

Alternatively, we introduce the \emph{upper convected deformation derivative} $ \gaugeU[\Wb]: \tangentS \rightarrow \tangentS $ by
\begin{align}
    \gaugeU[\Wb]\qb[\para]
        &:= ( \gauge{\Wb} q^i[\para] )\partial_i\para
        = \gaugeS[\Wb]\qb[\para]  - \Gb[\para,\Wb]\qb[\para]\formComma
    \label{eq:upper_deformation_derivativ_vector}
\end{align}
which acts on the contravariant proxy components with respect to the parametrization basis but is indeed basis independent as per the second equality following \eqref{eq:deformation_derivativ_vector}; and we define the following surface gauge:
\begin{definition}[Upper-Convected Gauge of Surface Independence on Tangential Vector Fields]\label{def:upper_goi_vector}
    A tangential vector field $ \qb[\para]\in \tangentS $ fulfills the gauge of upper convected surface independence $ \gaugeU\qb=0 $, if and only if
    one of the following equivalent statements is true:
    \begin{enumerate}[(i)]
    \item 
    $ \gaugeU[\Wb] \qb[\para] = 0 $ for all $ \Wb\in\tangentR $.
    \item Scalar gauges $ \gauge{}q^i=0 $ are fulfilled componentwise for the contravariant proxy field of $ \qb[\para] $.
    \end{enumerate}
\end{definition}

In comparison, we define a deformation derivative $ \gaugeL[\Wb]: \tangentS \rightarrow \tangentS $, 
which acts on the covariant proxies $ \{q_i[\para]\} $ with respect to the parametrization basis of a tangential vector field $ \qb[\para] $.
Following \eqref{eq:deformation_of_metric}, we find
\begin{align*}
    \gauge{\Wb} q_i[\para] 
        &= \gauge{\Wb} (g_{ij}[\para]q^j[\para])
         =  g_{ij}[\para]  \gauge{\Wb} q^j[\para] + 2S_{ij}[\para,\Wb]q^j[\para]
        \formPeriod
\end{align*}
We therefore may define the \emph{lower-convected deformation derivative} $ \gaugeL[\Wb]: \tangentS \rightarrow \tangentS $ as follows
\begin{align}
    \gaugeL[\Wb]\qb[\para]
        &:= g^{ij}[\para]( \gauge{\Wb} q_i[\para] )\partial_j\para
         = \gaugeS[\Wb]\qb[\para]  +  \Gb^T[\para,\Wb]\qb[\para]
    \label{eq:lower_deformation_derivativ_vector}
\end{align}
where we note that the definition is basis independant. Using the above, we define another surface gauge:
\begin{definition}[Lower-Convected Gauge of Surface Independence on Tangential Vector Fields]\label{def:lower_goi_vector}
    A tangential vector field $ \qb[\para]\in \tangentS $ fulfill the gauge of lower-convected surface independence $ \gaugeL\qb=0 $, if and only if
    one of the following equivalent statements is true:
    \begin{enumerate}[(i)]
    \item $ \gaugeL[\Wb] \qb[\para] = 0 $ is valid for all $ \Wb\in\tangentR $.
    \item Scalar gauges $ \gauge{}q_i=0 $ are fulfilled componentwise for the covariant proxy field of $ \qb[\para] $.
    \end{enumerate}
\end{definition}

Finally, we introduce another deformation derivative, which also applies on vector fields directly 
and affects corotational variations of the surface.
We take the corotational infinitesimal variation from \cite{SalbreuxJuelicherProstCallan-Jones__2022}.
In our notations, it is defined  by
\begin{align}
    \gaugeJ[\Wb]\qb[\para] := \gaugeS[\Wb]\qb[\para] - \Ab[\para,\Wb]\qb[\para] \formComma
    \label{eq:jaumann_deformation_derivativ_vector}
\end{align}
where $ 2\Ab[\para,\Wb] = (\Gb - \Gb^{T})[\para,\Wb] = \nabla \wb[\para] - (\nabla \wb[\para])^T $, see section \ref{sec:gradient_of_W} for more details.
We call the derivative $ \gaugeJ[\Wb]: \tangentS \rightarrow \tangentS $ 
\emph{Jaumann deformation derivative} of tangential vector fields in direction of $ \Wb\in\tangentR $.
The name prefix ``Jaumann'' is chosen consistently to the naming of time derivatives in  \cite{Nitschke_2022},
since time and deformation derivatives are structurally closely related.
For readers, who are used to other calculi or notations,
the term $ \Ab[\para,\Wb]\qb[\para] $
could also be written as $ \frac{1}{2}\proj_{\tangentS}\left((\nabla_{\R^3} \times \Wb) \times \qb[\para]\right) $, with the usual gradient and cross-product in $ \R^3 $,
or $ \frac{1}{2}(* d \wb[\para]^{\flat})(*\qb[x]^{\flat})^{\sharp} $ in context of differential forms, with Hodge- and musical isomorphisms and exterior derivative.
Comparing the Jaumann \eqref{eq:jaumann_deformation_derivativ_vector} with the upper-convected \eqref{eq:upper_deformation_derivativ_vector} and
the lower-convected deformation derivative \eqref{eq:lower_deformation_derivativ_vector} yields 
\begin{align*}
    \gaugeJ[\Wb] = \frac{1}{2}\left( \gaugeU[\Wb] + \gaugeL[\Wb] \right)
                :\ \tangentS \rightarrow \tangentS \formPeriod
\end{align*}
\begin{definition}[Jaumann Gauge of Surface Independence on Tangential Vector Fields]\label{def:jauman_goi_vector}
    A tangential vector field $ \qb[\para]\in \tangentS $ fulfill the gauge of Jaumann surface independence $ \gaugeJ\qb=0 $, if and only if
    one of the following equivalent statements is true:
    \begin{enumerate}[(i)]
    \item $ \gaugeJ[\Wb] \qb[\para] = 0 $ is valid for all $ \Wb\in\tangentR $.
    \item $ \gauge{\Wb}q^i[\para] + g^{ij}[\para]\gauge{\Wb}q_j[\para] = 0 $ is valid componentwise for the contra- and covariant proxy field of $ \qb[\para] $ and for  all $ \Wb\in\tangentR $.
    \end{enumerate}
\end{definition}

\begin{figure}
\centering
\includegraphics[width=.95\linewidth]{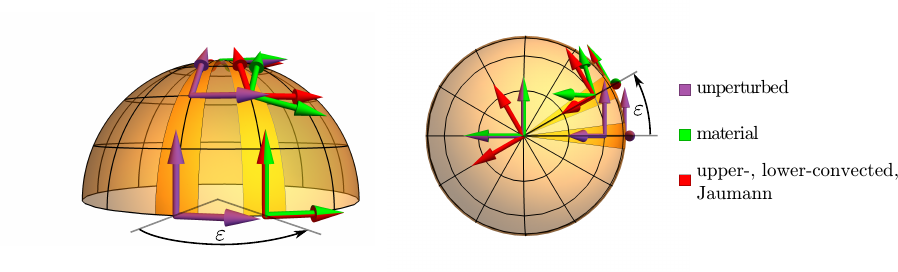}       
\caption{Rigid rotational deformation of a hemisphere (side and top view): 
           If $ \para $ parameterizes the hemisphere $ \surf $, 
           then 
           $\para_\eps = \para + \eps\Wb + \landau(\eps^2) $ parameterizes the rotated hemisphere $ \surf_{\eps}= \surf_{\eps\Wb} + \landau(\eps^2) $ by an angle $ \eps $.
           Two orthogonal tangential vector fields (\textcolor{lighter_purple_mathematica}{purple}) of equal lengths are shown on a fixed longitude and three different latitudes
           at $ \surf $.
           The \textcolor{green}{green} and \textcolor{red}{red} tangential vector fields are given on $ \surf_{\eps} $ by pushing-forward the purple ones 
           with the map $ \surf \rightarrow \surf_{\eps} $ assuming different gauges of surface independence, respectively.
           At the equator, where the axis of rotation lays parallel to the tangential plane of $ \surf $, all four gauges imply the same pushforward, 
           since the surface gradient of $ \Wb $ is vanishing, \cf\ table \ref{tab:gauge_deformation_derivative_vector}.
           At all other locations, only the symmetric gradient of $ \Wb $ is vanishing due to rigid body deformation, 
           \ie\ the upper-, lower-convected and Jaumann gauge yield equal pushforwards, whereas the material gauge implies a different one. 
           }
\label{fig:rot_sphere_vector}
\end{figure}

\begin{figure}
\centering
\includegraphics[width=.95\linewidth]{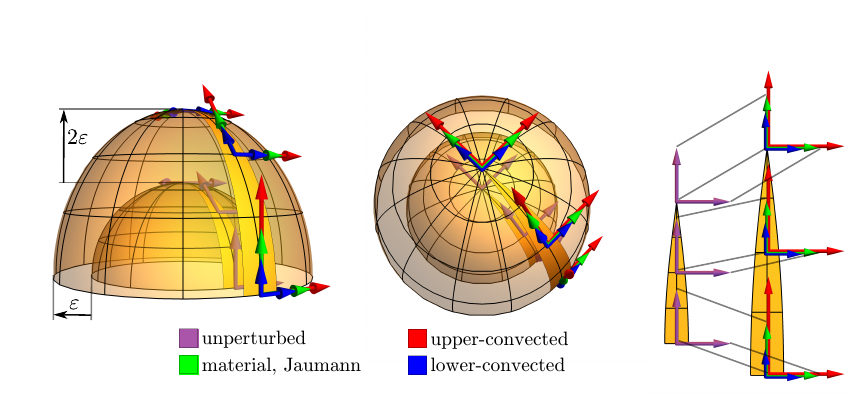}        
\caption{Strain deformation of a hemisphere (side and top view): 
           If $ \para $ parameterize the hemisphere $ \surf $, 
           then, for $ \Wb = \normal + \eb_z $, $ \para_\eps = \para + \eps\Wb$ parameterize the strained hemispheroid $\surf_{\eps\Wb} $.
           Two orthogonal tangential vector fields (\textcolor{lighter_purple_mathematica}{purple}) of equal lengths are shown on a fixed longitude and three different latitudes
           at $ \surf $ exactly like in figure \ref{fig:rot_sphere_vector}.
           The \textcolor{green}{green}, \textcolor{red}{red} and \textcolor{blue}{blue} tangential vector fields are given on $ \surf_{\eps} $ by pushing-forward the purple ones 
           with the map $ \surf \rightarrow \surf_{\eps} $ assuming different gauges of surface independence, respectively.
           The material and Jaumann gauge imply the same pushforward, since this is a pure strain deformation, 
           \ie\ the surface gradient of $ \Wb $ is symmetric, \cf\ table \ref{tab:gauge_deformation_derivative_vector}.
           Gray parallel lines on the unrolled arc-length segment (right) indicate length preserving for pushforwarded vectors due to material and Jaumann gauges.
           By contrast, pushforwarded vectors, \wrt\ the upper- or lower-convected gauge, are stretched along or against the deformation.
           However, the geometric mean of both yields the length preserved green vectors.}
\label{fig:stretch_sphere_vector}
\end{figure}

The four introduced gauges above differ from each other. 
However, a single assumed gauge completely determines all four deformation derivatives.
Table \ref{tab:gauge_deformation_derivative_vector} compares deformation derivatives under various assumptions of gauges of surface independence.
We observe that if a gauge and a deformation derivative belong to different types, the derivative takes the form $ \Psib[\para,\Vb]\qb[\para] $, 
where $\Psib[\para,\Vb]$ is either $ \Gb[\para, \Wb] = \nabla\wb[\para] - \wnor[\para]\shop[\para] $, its transpose, its symmetric part, or its skew-symmetric part as defined in section \ref{sec:gradient_of_W}.
As shown in table \ref{tab:gauge_deformation_derivative_vector}, if $\Wb$ is restricted to a subspace of $ \tangentR $, some of the various gauges can become equal.
For instance, if $ \Wb $ describes rigid deformations of the surface, then the upper-convected, lower-convected and Jaumann gauges would be equal, as illustrated in figure \ref{fig:rot_sphere_vector}.
However, if $ \Wb $ represents a pure strain deformation, the material and the Jaumann gauge would be the same,  as depicted in figure \ref{fig:stretch_sphere_vector}.

\begin{table}
\centering
\renewcommand{\arraystretch}{1.3}
\begin{tabular}{|r|c|c|c|c|}
\hline
\multirow{2}{*}{\diagbox{Derivs.}{Gauges}}  & $ \gaugeS\qb=0 $ &  $ \gaugeU\qb=0 $ & $ \gaugeL\qb=0 $ & $ \gaugeJ\qb=0 $ \\
& (Definition \ref{def:goi_vector}) & (Definition \ref{def:upper_goi_vector}) & (Definition \ref{def:lower_goi_vector}) & (Definition \ref{def:jauman_goi_vector})\\
\hline
\eqref{eq:deformation_derivativ_vector} $\;\;\; \gaugeS[\Wb]\qb[\para] $ & $ 0 $ & $ \Gb\qb[\para] $ & $ -\Gb^T\qb[\para] $ & $ \Ab\qb[\para] $\\
\hline
\eqref{eq:upper_deformation_derivativ_vector} $\;\;\; \gaugeU[\Wb]\qb[\para] $ & $ -\Gb\qb[\para] $ & $ 0 $ & $ -2\Sb\qb[\para] $ & $ -\Sb\qb[\para] $\\
\hline
\eqref{eq:lower_deformation_derivativ_vector} $\;\;\; \gaugeL[\Wb]\qb[\para] $ & $ \Gb^T\qb[\para] $ & $ 2\Sb\qb[\para] $ & $ 0 $ & $ \Sb\qb[\para] $\\
\hline
\eqref{eq:jaumann_deformation_derivativ_vector} $\;\;\; \gaugeJ[\Wb]\qb[\para] $ & $ -\Ab\qb[\para] $ & $ \Sb\qb[\para] $ & $ -\Sb\qb[\para] $ & $ 0 $\\
\hline
\end{tabular}
\caption{Deformation derivatives under various gauges of surface independence on tangential vector fields:
         Rows from top to bottom exhibit material, upper-convected, lower-convected and Jaumann deformation derivatives of a tangential vector field $ \qb[\para]\in\tangentS $.
         Columns from left to right indicate the assumed gauge of material, upper-convected, lower-convected or Jaumann surface independence.
         $ \Wb\in\tangentR $ is an arbitrary deformation field of the surface $ \surf $.
         We abbreviate the surface gradient of $ \Wb $ by
         $ \Gb := \Gb[\para,\Wb] = \nabla\wb[\para] - \wnor[\para]\shop[\para] $, see section \ref{sec:gradient_of_W}.
         The symmetric part of $ \Gb $ is $ \Sb := \Sb[\para,\Wb] $ and the antisymmetric part is $ \Ab := \Ab[\para,\Wb] $, 
         \cf\ table \ref{tab:gradients_of_W}.}
\label{tab:gauge_deformation_derivative_vector}
\end{table}

\subsubsection{Consequences for Energy Variations}\label{sec:consequences_for_energy_variations_vector}

For a given state comprising the tangential vector field $ \qb[\para] $ and a parameterization $ \para $, which fully determines the surface $ \surf $, 
the state energy $ \energy[\para, \qb[\para]]\in\R $. This allows to define directional total variations of $ \energy $ by
\begin{align}
    \inner[4]{\hilbertR}{\frac{\del\energy}{\del\para} , \Wb} \hspace{-0pt}
        &:= \lim_{\eps\rightarrow 0 } \frac{\energy[\para + \eps\Wb, \qb[\para + \eps\Wb]] - \energy[\para, \qb[\para]]}{\eps} \in\R \formComma
\label{eq:variation_X_raw_vector}\\
   \inner[4]{\hilbertS}{\frac{\del\energy}{\del\qb},  \rb} \hspace{-0pt}
        &:= \lim_{\eps\rightarrow 0 } \frac{\energy[\para, \qb[\para] + \eps\rb] - \energy[\para, \qb[\para]]}{\eps} \in\R
\label{eq:variation_q_raw_vector}
\end{align}
in arbitrary directions of $ \Wb\in\tangentR $ and $ \rb\in\tangentS $.
Variation \eqref{eq:variation_X_raw_vector} bears a subtle issue.
The term $ \energy[\para + \eps\Wb, \qb[\para + \eps\Wb]] $ is not determined, since  $ \qb[\para + \eps\Wb]\in\tangent{\surf_{\eps\Wb}} $ is not determined.
However, a Taylor expansion at $ \eps=0 $, \wrt\ the embedding space shows
\begin{align*}
    \qb[\para + \eps\Wb] 
            &= \proj_{\tangent{\surf_{\eps\Wb}}}\qb[\para + \eps\Wb]\\
            &=  \proj_{\tangent{\surf_{\eps\Wb}}}\left(\left(\qb[\para]\right)^{*_{\eps\Wb}} + \eps\left(\frac{d}{d\eps}\Big\vert_{\eps=0} \qb[\para + \eps\Wb]\right)^{\!\!*_{\eps\Wb}} + \landau(\eps^2)\right)
\end{align*}
where $ *_{\eps\Wb}:\tangentR \rightarrow \tangent{\R^3\vert_{\surf_{\eps\Wb}}} $ is the simplest conceivable $ \R^3 $-pushforward for the map $\surf \rightarrow \surf_{\eps\Wb}  $ 
and is given by
\begin{align}
    \Qb^{*_{\eps\Wb}} (y^1,y^2)
        &:= Q^{I}(y^1,y^2)\eb_I \in \tangent[_{(\para+\eps\Wb)(y^1,y^2)}]{\R^3}
     \label{eq:pushforward_vector_R3}
\end{align}
for all vector fields $ \Qb\in\tangentR $.
The deformation derivative $ \gauge{\Wb}: \tangentS \rightarrow \tangentR $, defined in \eqref{eq:deformation_derivativ_full_vector},
and another partial Taylor expansion at $ \proj_{\tangent{\surf_{\eps\Wb}}}\left(\qb[\para]\right)^{*_{\eps\Wb}}\in\tangent{\surf_{\eps\Wb}} $ of the energy $ \energy $ \wrt\ to the deformed surface $ \surf_{\eps\Wb} $ gives
\begin{align}
    \energy[\para + \eps\Wb, \qb[\para + \eps\Wb]] \hspace{-90pt}\notag\\
        &= \energy[\para + \eps\Wb, \proj_{\tangent{\surf_{\eps\Wb}}}\left(\qb[\para]\right)^{*_{\eps\Wb}}
                                     + \eps \proj_{\tangent{\surf_{\eps\Wb}}}\left(\gauge{\Wb}\qb[\para]\right)^{*_{\eps\Wb}} + \landau(\eps^2)]\notag\\
        &=  \energy[\para + \eps\Wb, \proj_{\tangent{\surf_{\eps\Wb}}}\left(\qb[\para]\right)^{*_{\eps\Wb}}]
    \label{eq:taylor_energy_at_X}\\
        &\quad\quad    + \eps \inner{\hilbertT{\surf_{\eps\Wb}}}{\frac{\del\energy}{\del(\proj_{\tangent{\surf_{\eps\Wb}}}\qb^{*_{\eps\Wb}})} 
                                                        , \proj_{\tangent{\surf_{\eps\Wb}}}\left(\gauge{\Wb}\qb[\para]\right)^{*_{\eps\Wb}}}
            + \landau(\eps^2)\formComma \notag
\end{align}
where we used the directional variation \eqref{eq:variation_q_raw_vector}, \wrt\ tangential vector field 
$ \proj_{\tangent{\surf_{\eps\Wb}}}\left(\qb[\para]\right)^{*_{\eps\Wb}}\in\tangent{\surf_{\eps\Wb}} $
in direction of $ \proj_{\tangent{\surf_{\eps\Wb}}}\left(\gauge{\Wb}\qb[\para]\right)^{*_{\eps\Wb}}\in\tangent{\surf_{\eps\Wb}} $,
and properties of inner products.
To relate the total variation \eqref{eq:variation_X_raw_vector}, we define the partial variation
\begin{align}
    \inner[4]{\hilbertR}{\frac{\partial\energy}{\partial\para},  \Wb} \hspace{-10pt}
            &:= \lim_{\eps\rightarrow 0 } \frac{\energy[\para + \eps\Wb, \proj_{\tangent{\surf_{\eps\Wb}}}\left(\qb[\para]\right)^{*_{\eps\Wb}}] - \energy[\para, \qb[\para]]}{\eps} \in\R
   \label{eq:variation_X_partial_raw_vector}
\end{align}
in arbitrary directions of $ \Wb\in\tangentR $ \wrt\ $ \hilbertR $.
In practice, this derivative means that everything in $ \energy $ has to be differentiated \wrt\ $ \para $ except for $ \qb[\para]\in\tangentS$ (incl.\ its frame).
For instance, in the energy term $ \int_{\surf} \|\qb[\para]\|^2_{\tangentS} \mu[\para] $ only $ \mu[\para] $ has to be differentiated.
Using that $ \proj_{\tangent{\surf_{\eps\Wb}}} \circ *_{\eps\Wb} \rightarrow \proj_{\tangentS} $ is valid for $ \eps\rightarrow 0 $,
the material deformation derivative \eqref{eq:deformation_derivativ_vector}
and substituting the Taylor expansion above into the total variation \eqref{eq:variation_X_raw_vector} yields
\begin{align}
    \inner[4]{\hilbertR}{\frac{\del\energy}{\del\para} , \Wb}
        &= \inner[4]{\hilbertR}{\frac{\partial\energy}{\partial\para} , \Wb}
            + \inner[4]{\hilbertS}{ \frac{\del\energy}{\del\qb},  \gauge[\surf]{\Wb}\qb[\para] } \formPeriod
     \label{eq:deformation_transport_vector}
\end{align}
Note that the material deformation derivative $ \gauge[\surf]{\Wb}\qb[\para]\in\tangentS $ in \eqref{eq:deformation_transport_vector} depends on the chosen gauge of surface independence
and can be obtained from the first row of table \ref{tab:gauge_deformation_derivative_vector}.

Some readers may be used to consider rather contravariant proxys of vector fields than an entire field with its frame.
Hence, we give with \eqref{eq:deformation_transport_vector_contarvariantly} in appendix \ref{sec:upper_deformation_transport_vector} an equivalent identity
\wrt\ to an energy $ \tenergy[\para,q^1[\para],q^2[\para]] $, 
but in terms of a contravariant approach.
Also treating the proxy of the metric tensor explicitly could be of advantage.
For this purpose \eqref{eq:deformation_transport_vector_with_metric} in appendix \ref{sec:deformation_transport_vector_with_metric} gives an equivalent identity
\wrt\ an energy $ \henergy[\para, \qb[\para], \{g_{ij}[\para]\}] $.
We like to highlight that $ \inner{\hilbertR}{\frac{\del\energy}{\del\para} , \Wb} $ depends only on the semantical description of $ \energy $ and the assumed gauge of independence.
Even if we write the same energy in dependency of $ [\para, \{q^i[\para]\}, \{q_i[\para]\}, \qb[\para],\{g^{ij}[\para]\}, \{g_{ij}[\para]\}, \shop[\para], \meanc[\para], \gaussc[\para], \ldots] $
the result is still the same, 
which shows the advantage of total derivatives.
This is also the reason to omit geometrical quantities other than $ \para $ in the dependency list, since all of them are represented by $ \para $ and would have the only effect to blow up
formulas without an influence on the results.    

Sometimes it could be difficult to calculate the partial variation of an energy without touching its tangential vector field argument.
Or it is hard to obtain the total variation for the chosen gauge of surface independence, though it is much easier for another gauge.
In such situations it could be helpful to use the identities
\begin{align}
    \inner[4]{\hilbertR}{\frac{\del\energy}{\del\para} , \Wb} \hspace{-10pt}
        &= \inner[4]{\hilbertR}{\frac{\del\energy}{\del\para}\Big\vert_{\gauge[\Psi]{}\qb=0} , \Wb}
            \hspace{-15pt}+ \inner[4]{\hilbertS}{ \frac{\del\energy}{\del\qb},  \gauge[\Psi]{\Wb}\qb[\para] }
      \label{eq:deformation_change_vector} \\
    \inner[4]{\hilbertR}{\frac{\partial\energy}{\partial\para} , \Wb} \hspace{-10pt}
        &= \inner[4]{\hilbertR}{\frac{\del\energy}{\del\para}\Big\vert_{\gauge[\Psi]{}\qb=0} , \Wb}
           \hspace{-15pt}+ \inner[4]{\hilbertS}{ \frac{\del\energy}{\del\qb},  \gauge[\Psi]{\Wb}\qb[\para] - \gauge[\surf]{\Wb}\qb[\para] },
      \label{eq:deformation_change_vector_partial}
\end{align}
where the symbol $ \Psi\in\{\surf,\sharp,\flat,\jau\} $ determine the desired deformation derivative.
We get both identities by using \eqref{eq:deformation_transport_vector}  and the fact that the partial variation 
\eqref{eq:variation_X_partial_raw_vector} does not depend on any gauge.
Moreover, \eqref{eq:deformation_change_vector} generalizes \eqref{eq:deformation_transport_vector}
in the sense that \eqref{eq:deformation_change_vector} equals \eqref{eq:deformation_transport_vector} for $ \Psi=\surf $, since
$ \frac{\del\energy}{\del\para}\big\vert_{\gaugeS{}\qb=0} = \frac{\partial\energy}{\partial\para} $. 

\subsubsection{Consequences for $ \hil $-Gradients} \label{sec:L2_gradient_vector}
    The $ \hil $-gradient is weakly given by
    \begin{align*}
        \innerH{\tangentR \times \tangentS}{ \hilgrad\energy, ( \Wb, \rb ) }
            &:= \inner[4]{\hilbertR}{\frac{\del\energy}{\del\para} , \Wb}
              + \inner[4]{\hilbertS}{\frac{\del\energy}{\del\qb} , \rb}
    \end{align*}
    for all $( \Wb, \rb )\in\tangentR \times \tangentS$.
    Considering \eqref{eq:deformation_transport_vector} and a chosen gauge of independence for $ \qb[\para]\in\tangentS $, 
    we get
    \begin{align}
        &\innerH{\tangentR \times \tangentS}{ \hilgrad\energy, ( \Wb, \rb ) } \notag\\
        &\quad\quad= \inner[4]{\hilbertR}{\frac{\partial\energy}{\partial\para} , \Wb}
              + \inner[4]{\hilbertS}{\frac{\del\energy}{\del\qb} , \rb + \Psib[\para,\Wb]\qb[\para]}\formComma
        \label{eq:L2Grad_weak}
    \end{align}
    \begin{center}
    \begin{tabular}{lr|c|c|c|cl}
        where &$ \Psib = $& $0$ & $\Gb$ & $ -\Gb^T $ & $ \Ab $\\
        for & $0=$ & $ \gaugeS\qb $ &  $ \gaugeU\qb $ & $ \gaugeL\qb $ & $ \gaugeJ\qb $ &
    \end{tabular}
    \end{center}
    according to the first row in table \ref{tab:gauge_deformation_derivative_vector}.
    The strong formulation of the $ \hil $-gradient reads
    \begin{align}
        \hilgrad\energy 
            &= \left( \frac{\del\energy}{\del\para}, \frac{\del\energy}{\del\qb} \right) 
             = \left( \frac{\partial\energy}{\partial\para} - \div_{\tangentS[^2]}\gstress, \frac{\del\energy}{\del\qb}  \right) \in \tangentR \times \tangentS \formComma
          \label{eq:L2Grad_strong}
    \end{align}
    where we used \eqref{eq:L2Grad_weak}, the deformation-$ \hil $-adjoints $ \Psib^{\dadj}[\para] $ given in \eqref{eq:deformation_adjoints}
    and  the gauge stress tensor fields $ \gstress\in\tangentS[^2] $, which are depending on the chosen gauge and given in the first row of table \ref{tab:gauge_stress}.
    In an equation of motion containing $ \hilgrad\energy $, \eg\ in the  $ \hil $-gradient flow in the subsection below, 
    the term $ \div_{\tangentS[^2]}\gstress = \div\gstress + \left\langle \shop[\para],\gstress \right\rangle\normal\in\tangentR $ can be interpreted
    as a partial force induced by the chosen gauge of surface independence. 
    We name the resulting force ``partial'', 
    since $ \gstress $ depends on the syntactical description of $ \energy $.
    For instance, if we use an equivalent contravariant approach $ \tenergy[\para,q^1[\para],q^2[\para]]= \energy[\para,\qb[\para]] $,
    as given in appendix \ref{sec:upper_deformation_transport_vector},
    we see in \eqref{eq:L2Grad_strong_contravariant_approach} that we end up with a gauge stress field 
    $ \tgstress = \gstress - \frac{\del\energy}{\del\qb}\otimes\qb[\para]  \in\tangentS[^2] $, see second row in table \ref{tab:gauge_stress}.
    Note that the approach in appendix \ref{sec:deformation_transport_vector_with_metric}, 
    where we treat the metric proxy explicitly by $ \henergy[\para, \qb[\para], \{g_{ij}[\para]\}]=\energy[\para,\qb[\para]] $
    and using the term $ \frac{\partial\henergy}{\partial\para} $ instead $ \frac{\partial\energy}{\partial\para} $,
    yields the stress tensor fields $ \hgstress = \gstress+ \frac{\partial\henergy}{\partial g_{ij}}\partial_i\para\otimes\partial_j\para $.
    \begin{table}
    \centering
    \renewcommand{\arraystretch}{1.5}
    \begin{tabular}{|r|c|c|c|c|}
    \hline
        & $ \gaugeS\qb=0 $ 
            &  $ \gaugeU\qb=0 $ 
                & $ \gaugeL\qb=0 $ 
                    & $ \gaugeJ\qb=0 $ \\
        & (Definition \ref{def:goi_vector}) 
            & (Definition \ref{def:upper_goi_vector}) 
                & (Definition \ref{def:lower_goi_vector}) 
                    & (Definition \ref{def:jauman_goi_vector})\\
    \hline
    $ \gstress $ 
        & $ 0 $ 
            & $ \frac{\del\energy}{\del\qb}\otimes\qb[\para] $ 
                & $ -\qb[\para]\otimes\frac{\del\energy}{\del\qb} $ 
                    & $ \frac{\del\energy}{\del\qb} \owedge \qb[\para] $\\
    \hline
    $ \tgstress $ 
        & $ -\frac{\del\tenergy}{\del\qb}\otimes\qb[\para] $ 
            & $ 0 $ 
                & $ -2 \frac{\del\tenergy}{\del\qb} \ovee \qb[\para] $ 
                    & $ -\frac{\del\tenergy}{\del\qb} \ovee \qb[\para] $\\
    \hline
    \end{tabular}
    \caption{Gauge stress tensor fields $ \gstress,\tgstress\in\tangentS[^2] $ caused by different gauges of surface independence.
                $ \gstress $, \resp\ $ \tgstress $, apply in \eqref{eq:L2Grad_strong} and \eqref{eq:FOModel}, \resp\ \eqref{eq:L2Grad_strong_contravariant_approach},  
                according to syntactically stipulated partial dependency in the energy $ \energy $, or $\tenergy$.
                We define anti- and symmetric outer products by 
                $ 2\rb\owedge\qb := \rb\otimes\qb  - \qb\otimes\rb $ and $ 2\rb\ovee\qb := \rb\otimes\qb  + \qb\otimes\rb $ for
                all $ \rb,\qb\in\tangentS $.}
    \label{tab:gauge_stress}
    \end{table}
    
    \begin{remark}
        The solutions of the stationary equation $ 0=\hilgrad\energy $ do not depend on any gauge of surface independence. On the one hand $ \frac{\partial\energy}{\partial\para} $ does not depend on a gauge.
        On the other hand at a local extremum of $ \energy $, where $ \frac{\del\energy}{\del\qb}=0 $ is valid, all gauge stress tensor fields $ \gstress $ are vanishing.
    \end{remark}

    \subsubsection{Consequences for $ \hil $-Gradient Flows}\label{sec:consequences_gradient_flow_vector}   
    
    In general, the  strong formulation of the $ \hil $-gradient flow reads
    \begin{align}
        \left( \frac{d}{dt}\para, \timeD\qb[\para] \right)
            &= -\lambda \hilgrad\energy  \formComma
        \label{eq:L2flow}
    \end{align}
    where $ \lambda > 0$ is a mobility parameter coefficient and 
    $ \timeD\qb[\para] \in \tangentS $
    an observer-invariant instantaneous time derivative of $ \qb[\para]\in\tangentS  $  given in \cite[Conclusion 6]{Nitschke_2022}.
    We consider the material, upper-convected, lower-convected and Jaumann time derivative in this context. 
    At this, the parameterization $ \para $ is time-depending, as well as the tangential vector field $ \qb[\para]\in \tangentS $
    by $ \para $ and also partially in $ t $, 
    \ie\ at an event $ (t,\para\vert_{t}(y^1,y^2))\in\R\times\surf\vert_{t}\subset\R^4 $ describes  
    $ \para(t,y^1,y^2) := \para\vert_{t}(y^1,y^2)\in\surf\vert_{t} $ the spatial location and 
    $ \qb[\para](t,y^1,y^2)=(\qb[\para](y^1,y^2))\vert_t\in\tangent[_{\para\vert_{t}(y^1,y^2)}]{\surf\vert_{t}} $ the state of the vector field.
    We assume that the energy $ \energy $ is instantaneous and not partially time depending,
    \ie\ $ \energy $ does not depend on any rates of $ \para $ or $ \qb[\para] $ and it holds $(\energy[\para,\qb[\para]])\vert_{t} = \energy[\para\vert_{t},(\qb[\para])\vert_{t}]$,
    which is common for potential or free energies.
    We omit the affix $ \vert_{t} $ if it is clear that quantities are given at time $ t $.
    
    Note that the time depending parameterization $\para$ can be any observer parameterization for the time depending surface $\surf$.
    This can also be the material parameterization $\para_\mfrak$.
    By fixing a single material particle at $ (y^1_\mfrak,y^2_\mfrak) $ in the parameter space,
    $\para_\mfrak\vert_{(y^1_\mfrak,y^2_\mfrak)}(t)$ describes
    the world line of this material particle. 
    While the material parameterization takes the Lagrange perspective, 
    the observer parameterization $\para_\ofrak$ is an arbitrary parameterization taking the observer perspective,
    \ie\ $\para_\ofrak\vert_{(y^1_\ofrak,y^2_\ofrak)}(t)$ describes the world line of an observer particle, see \cite{Nitschke_2022}.
    The left-hand side of \eqref{eq:L2flow} depends on the dynamics of the material, but we are able to formulate it invariantly of the observer.
    The first component is the material velocity field
    $\Vb_{\mfrak}[\para_\mfrak] := \frac{d}{dt}\para_\ofrak = \partial_t\para_\mfrak \in\tangentR$.
    Whereas $\Vb_\ofrak[\para_\ofrak]:=\partial_t\para_\ofrak\in\tangentR$ is the observer velocity field.
    From observer perspective, both velocity fields can be related by the relative velocity
    $\ub[\para_\ofrak,\para_\mfrak](t,y^1_\ofrak,y^1_\ofrak)  
        := \Vb_{\mfrak}^{\ofrak}[\para_\ofrak,\para_\mfrak](t,y^1_\ofrak,y^1_\ofrak) - \Vb_\ofrak[\para_\ofrak](t,y^1_\ofrak,y^1_\ofrak)$,
    where $\Vb_{\mfrak}^{\ofrak}[\para_\ofrak,\para_\mfrak](t,y^1_\ofrak,y^1_\ofrak) 
                := \Vb_{\mfrak}[\para_\mfrak](t,(\para_\mfrak\vert_t)^{-1}(\para_\ofrak(t,y^1_\ofrak,y^1_\ofrak)))$
    is the material velocity evaluated at observer location $\para_\ofrak(t,y^1_\ofrak,y^1_\ofrak)$.
    Note that for the relative velocity field holds $\ub[\para_\ofrak,\para_\mfrak]\in\tangentS$, since the normal part of the observer and material velocity is equal. 
    Therefore, the  material time derivative of $ \qb[\para_\ofrak]\in\tangentS $ is given by 
    \begin{align*}
        \dot{\qb}[\para_\ofrak] 
            &= (\partial_t q^{i}[\para_\ofrak])\partial_i\para_\ofrak + \nabla_{\ub[\para_\ofrak,\para_\mfrak]}\qb[\para_\ofrak]  + \Gb[\para_\ofrak,\Vb_\ofrak[\para_\ofrak]]\qb[\para_\ofrak]\\
            &= \proj_{\tangentS} \frac{d}{dt}\qb[\para_\ofrak] \formComma
    \end{align*}
    where the rear identity is a conclusion in \cite[Proposition 4]{Nitschke_2022}.
    All considered time derivatives can be derived from the material one by
    \begin{align}
        \timeD\qb[\para_\ofrak] &= \dot{\qb}[\para_\ofrak] - \Phib[\para_\ofrak,\Vb_{\mfrak}^{\ofrak}[\para_\ofrak,\para_\mfrak]]\qb[\para_\ofrak]\formComma
        \label{eq:time_derivatives_vector}
    \end{align}
    \begin{center}
    \begin{tabular}{lr|c|c|c|cl}
     where &$ \timeD\qb = $& $\dot{\qb}$ & $\timeLu\qb$ & $  \timeLl\qb $ & $ \timeJ\qb $   \\
     for &$ \Phib = $& $0$ & $\Gb$ & $ -\Gb^T $ & $ \Ab $ &,
    \end{tabular}
    \end{center}
    where $ \timeLu $ is the upper-convected, $ \timeLl $ the lower-convected and $ \timeJ $ the Jaumann time derivative on tangential vector fields, see \cite{Nitschke_2022}.
    
    Since the left-handed side of \eqref{eq:L2flow} is observer-invariant and the right-handed side is instantaneous,
    we take the Lagrangian perspective in the following  without loss of generality.
    This means that we set $ \para := \para_\ofrak = \para_\mfrak$.
    Due to this it is $\Vb[\para]:= \Vb_{\mfrak}[\para_\mfrak] = \Vb_{\mfrak}^{\ofrak}[\para_\ofrak,\para_\mfrak] = \Vb_\ofrak[\para_\ofrak]$ and 
    $\ub[\para_\ofrak,\para_\mfrak]=0$ valid.
    
    The time derivative of the energy $ \energy $ gives the energy rate
    \begin{align*}
        \frac{d}{dt}\energy
            &:= \lim_{\eps\rightarrow 0 } \frac{(\energy[\para,\qb[\para]])\vert_{t+\eps} - (\energy[\para,\qb[\para]])\vert_{t}}{\eps}\\
            &= \innerH{\tangentR}{\frac{\partial\energy}{\partial\para} , \Vb[\para]}
               + \innerH{\tangentS}{\frac{\del\energy}{\del\qb} , \dot{\qb}[\para]} \formPeriod
    \end{align*}
    This can be obtained by Taylor expansions at $ \epsilon=0 $, similarly to \eqref{eq:taylor_energy_at_X} for $ \Wb=\Vb[\para] + \landau(\eps) $. 
    A priori, the energy rate is invariant \wrt\ any chosen gauges of independence.
    This behavior changes if we substitute the gauge-dependent gradient flow \eqref{eq:L2flow} into this rate. 
    With \eqref{eq:deformation_transport_vector} we get
    \begin{align}
        \frac{d}{dt}\energy
            &= \innerH[2]{\tangentR}{\frac{\del\energy}{\del\para} , \Vb[\para]}
                           + \innerH[2]{\tangentS}{\frac{\del\energy}{\del\qb} , \dot{\qb}[\para] - \gauge[\surf]{\Vb[\para]}\qb[\para]}\notag\\
            &= \innerH[2]{\tangentR}{\frac{\del\energy}{\del\para} , \Vb[\para]}
                                       \hspace{-10pt}+ \innerH[2]{\tangentS}{\frac{\del\energy}{\del\qb} , \timeD\qb[\para] + (\Phib-\Psib)[\para,\Vb[\para]]\qb[\para]}
         \label{eq:dissipation_before_substituting_flow}\\
            &= - \lambda\left\| \hilgrad\energy \right\|_{\hil(\tangentR\times\tangentS)}^2
                 + \innerH[2]{\tangentS}{\frac{\del\energy}{\del\qb} , (\Phib-\Psib)[\para,\Vb[\para]]\qb[\para]} \formComma
         \label{eq:dissipation}
    \end{align}
    where $ \Psib[\para,\Vb[\para]]\in\tangentS[^2] $ is already introduced in \eqref{eq:L2Grad_weak} and depends on the chosen gauge,
    \cf\ the header in table \ref{tab:gauge_vs_time_derivative}.
    Using a time derivative in the gradient flow \eqref{eq:L2flow}, whose name is consistent with the chosen gauge, \eg\ upper-convected time derivative for a upper-convected gauge,
    is sufficient to ensure a dissipative energy, \ie\ it holds $ \frac{d}{dt}\energy \le 0 $, since $ (\Phib-\Psib)[\para,\Vb[\para]] = 0 $ is valid.
    In contrast, using time derivatives and gauges inconsistently in their names, \eg\ material time derivative and Jaumann gauge,
    could rise an issue for the dissipation, especially for small $ \lambda $.
    However, we present in table \ref{tab:gauge_vs_time_derivative} all 16 possible combinations we discussed.   
    \begin{table}
    \centering
    \renewcommand{\arraystretch}{1.3}
    \begin{tabular}{|>{\hspace{-3pt}}r<{\hspace{-3pt}}>{\hspace{-3pt}}l<{\hspace{-3pt}}|
                    >{\hspace{-3pt}}c<{\hspace{-3pt}}|
                        >{\hspace{-3pt}}c<{\hspace{-3pt}}|
                            >{\hspace{-3pt}}c<{\hspace{-3pt}}|
                                >{\hspace{-3pt}}c<{\hspace{-3pt}}|}
    \hline
    && $ \gaugeS\qb=0 $  
        &  $ \gaugeU\qb=0 $ 
            & $ \gaugeL\qb=0 $ 
                & $ \gaugeJ\qb=0 $ \\
    \multicolumn{2}{|c|}{$ (\Phib-\Psib)[\para,\Vb[\para]]$} 
    & (Definition\ref{def:goi_vector}) 
        & (Definition \ref{def:upper_goi_vector}) 
            & (Definition \ref{def:lower_goi_vector}) 
                & (Definition \ref{def:jauman_goi_vector})\\
    && $\Rightarrow\Psib=0$ 
        & $\Rightarrow\Psib=\Gb$ 
            & $\Rightarrow\Psib=-\Gb^T$ 
                & $\Rightarrow\Psib=\Ab$ \\
    \hline
    $\dot{\qb}[\para]$ & $ \Rightarrow \Phib=0$ 
    & $ 0 $ 
        & $ -\Gb[\para,\Vb[\para]] $ 
            & $ \Gb^T[\para,\Vb[\para]] $ 
                & $ -\Ab[\para,\Vb[\para]] $\\
    \hline
    $\timeLu\qb[\para]$ & $ \Rightarrow \Phib=\Gb$ 
    & $ \Gb[\para,\Vb[\para]] $ 
        & $ 0 $ 
            & $ 2\Sb[\para,\Vb[\para]] $ 
                & $ \Sb[\para,\Vb[\para]] $\\
    \hline
    $\timeLl\qb[\para]$ & $ \Rightarrow \Phib=-\Gb^{T}$ 
    & $ -\Gb^T[\para,\Vb[\para]]$ 
        & $ -2\Sb[\para,\Vb[\para]] $ 
            & $ 0 $ 
                & $ -\Sb[\para,\Vb[\para]] $\\
    \hline
    $\timeJ\qb[\para]$ & $ \Rightarrow \Phib=\Ab$ 
    & $ \Ab[\para,\Vb[\para]] $ 
        & $ -\Sb[\para,\Vb[\para]] $ 
            & $ \Sb[\para,\Vb[\para]] $ 
                & $ 0 $\\
    \hline
    \end{tabular}
    \caption{Determining $ (\Phib-\Psib)[\para,\Vb[\para]]\in\tangentS[^2] $ in the rear term of energy rate \eqref{eq:dissipation}
             by combinations of different chosen gauges of surface independence (columns) and time derivatives (rows).
             From left to right, \resp\ top to bottom, it is material, upper-convected, lower-convected, Jaumann gauge, 
             \resp\ time derivative, considered.
             For definitions of $ \Gb, \Sb $ and $ \Ab $ see table \ref{tab:gradients_of_W}
             and for time derivatives \cite[Conclusion 6]{Nitschke_2022}.
             Note that $ \Vb $ is the material as well as the observer velocity.}
    \label{tab:gauge_vs_time_derivative}
    \end{table}

   \begin{remark}\label{rem:onsager_vector}
    Using the Onsager's variational principle \cite{Doi} to obtain the gradient flow \eqref{eq:L2flow} 
    is forcing a consistent choice of time derivative and gauge of surface independence.
    This follows by minimizing the Rayleighian 
    \begin{align*}
        \rayleighian[\Vb,\timeD\qb] 
                &:= \frac{1}{2\lambda}\left( \left\| \Vb \right\|_{\hil(\tangentR)}^2  +  \left\| \timeD\qb \right\|_{\hil(\tangentR)}^2 \right) + (\frac{d}{dt}\energy)[\Vb,\timeD\qb]
    \end{align*}
    by the necessary condition $ (0,0) = (\frac{\del\rayleighian}{\del\Vb}, \frac{\del\rayleighian}{\del(\timeD\qb)}) $,
    with an energy rate $ \frac{d}{dt}\energy $ given in \eqref{eq:dissipation_before_substituting_flow} 
    \wrt\ process variables $ \Vb $ and $ \timeD\qb $.
    As a consequence, $ (\Phib-\Psib)[\para,\Vb[\para]] = 0 $ has to be valid by comparison with the gradient flow \eqref{eq:L2flow}. 
   \end{remark}
    
In section \ref{sec:example_vector} we provide an example for a Frank-Oseen-Helfrich energy, derive the $ \hil $-gradient flow and demonstrate differences of the consistent combinations in table \ref{tab:gauge_vs_time_derivative}.

\subsection{Tangential 2-Tensor Fields}\label{sec:tangential_tensor_fields}   

The proceeding is almost the same as for tangential vector fields in section \ref{sec:vector_fields}.
Therefore we keep the argumentation much shorter in this section.

\subsubsection{Gauges of Surface Independence}

A tangential 2-tensor field $ \qb[\para]\in\tangentS[^2] $ can be represented in a tangential as well as the Euclidean frame, 
\ie\ it holds $ \qb[\para] = q^{ij}[\para]\partial_i\para\otimes\partial_j\para = q^{IJ}[\para]\eb_I\otimes\eb_{J} $.
Since $ \eb_{I} $ does not depend on the surface in any way, we define and calculate
\begin{align}
    \gauge{\Wb} \qb[\para] 
           &:= \left(\gauge{\Wb} q^{IJ}[\para]\right) \eb_I \otimes\eb_{J} \notag\\
           &=  \left(\gauge{\Wb}q^{ij}[\para] \right) \partial_i\para\otimes\partial_j\para 
                +q^{ij}[\para]\left( \partial_i\Wb\otimes\partial_j\para  + \partial_i\para\otimes\partial_j\Wb  \right)
     \label{eq:deformation_derivative_tensor_raw}
\end{align}
\wrt\ the deformation derivative \eqref{eq:deformation_derivativ_scalar} on scalar fields.
The derivative \eqref{eq:deformation_derivative_tensor_raw} maps to $ \tangentR[^2] $, but all non-tangential parts of the image are not bearing any options for a possible gauge.
Therefore, we define the \emph{material deformation derivative}
$ \gaugeS[\Wb]: \tangentS[^2] \rightarrow \tangentS[^2] $ of tangential 2-tensor fields
in direction of $ \Wb\in\tangentR $ by
\begin{align} \label{eq:deformation_derivativ_tensor}
    \gaugeS[\Wb]\qb[\para] 
        &:=  (\proj_{\tangentS[^2]}[\para] \circ \gauge{\Wb})\qb[\para] \\
         &=  \left(\gauge{\Wb}q^{ij}[\para] \right) \partial_i\para\otimes\partial_j\para 
            +\Gb[\para,\Wb]\qb[\para]  + \qb[\para]\Gb^T[\para,\Wb] \formPeriod \notag 
\end{align}

\begin{definition}[Gauge of Material Surface Independence on Tangential 2-tensor Fields]\label{def:goi_tensor}
    A tangential tensor field $ \qb[\para]\in \tangentS[^2] $ fulfill the gauge of material surface independence $ \gaugeS\qb=0 $, if and only if
    $ \gaugeS[\Wb] \qb[\para] = 0 $ is valid for all $ \Wb\in\tangentR $.
\end{definition}

The \emph{$ \natural $-convected deformation derivatives} $ \gaugeN[\Wb]: \tangentS[^2] \rightarrow \tangentS[^2] $ of tangential 2-tensor fields
in direction of $ \Wb\in\tangentR $ are given in the four flavors $ \natural\in\{\sharp\sharp, \flat\flat, \sharp\flat, \flat\sharp\} $
and systematically named in this order by prefixes \emph{upper-upper} (or \emph{fully-upper}), \emph{lower-lower} (or \emph{fully-lower}), \emph{upper-lower} and \emph{lower-upper}.
They are principally affected by the scalar deformation of the $ \natural $-proxy matrix in $ (\tangentS[^0])^{2\times2} $ given by respective ``index-height'', 
\ie\ we define
\begin{align}
    \gaugeN[\Wb]\qb[\para] &=
    \left\{
    \begin{aligned}
    \gaugeUU[\Wb]\qb[\para]
        &:= \left(\gauge{\Wb}q^{ij}[\para] \right) \partial_i\para\otimes\partial_j\para\\
    \gaugeLL[\Wb]\qb[\para]
            &:= g^{ik}[\para]g^{jl}[\para]\left(\gauge{\Wb}q_{kl}[\para] \right) \partial_i\para\otimes\partial_j\para\\
    \gaugeUL[\Wb]\qb[\para]
                &:= g^{jk}[\para]\left(\gauge{\Wb}\tensor{q}{^i_k}[\para] \right) \partial_i\para\otimes\partial_j\para\\
    \gaugeLU[\Wb]\qb[\para]
                    &:= g^{ik}[\para]\left(\gauge{\Wb}\tensor{q}{_k^j}[\para] \right) \partial_i\para\otimes\partial_j\para
     \formPeriod
     \end{aligned}
     \right.
 \label{eq:deformation_derivative_components_natural_tensor}
\end{align}
Calculating the relation to the material deformation derivative results in
\begin{align}
    \gaugeN[\Wb]\qb[\para] 
        &=  \gaugeS[\Wb]\qb[\para] - \Psib_1[\para,\Wb]\qb[\para] - \qb[\para]\Psib_2^T[\para,\Wb] \formComma
    \label{eq:deformation_derivative_natural_tensor}
\end{align}
\begin{center}
\begin{tabular}{lr|c|c|c|cl}
 where &$ \Psib_1 = $& $\Gb$ & $-\Gb^T$ & $\Gb$ & $-\Gb^T$   \\
 and &$ \Psib_2 = $& $\Gb$ & $-\Gb^T$ & $-\Gb^T$ & $\Gb$   \\
 for &$ \natural = $& $\sharp\sharp$ & $\flat\flat$ & $ \sharp\flat $ & $ \flat\sharp $ &.
\end{tabular}
\end{center}

\begin{definition}[Gauge of $ \natural $-Convected Surface Independence on Tangential 2-Tensor Fields]\label{def:natural_goi_tensor}
    A tangential 2-tensor field $ \qb[\para]\in \tangentS[^2] $ fulfill the gauge of $ \natural $-convected surface independence $ \gaugeN\qb=0 $
    for $ \natural\in\{\sharp\sharp, \flat\flat, \sharp\flat, \flat\sharp\} $, if and only if
    one of the following equivalent statements is true:
    \begin{enumerate}[(i)]
    \item $ \gaugeN[\Wb] \qb[\para] = 0 $ is valid for all $ \Wb\in\tangentR $.
    \item Scalar gauges\\ 
        \begin{tabular}{lrl}
            & $ \gauge{} q^{ij}= 0 $ & for $ \natural=\sharp\sharp $,\\
            or & $ \gauge{} q_{ij}= 0 $ & for $ \natural=\flat\flat $,\\
            or & $ \gauge{} \tensor{q}{^i_j}= 0 $ & for $ \natural=\sharp\flat $,\\
            or & $ \gauge{} \tensor{q}{_i^j}= 0 $ & for $ \natural=\flat\sharp $,\\
        \end{tabular}\\
        are fulfilled componentwise for the respective proxy field of $ \qb[\para] $.
    \end{enumerate}
\end{definition}

We define the \emph{Jaumann deformation derivative} $ \gaugeJ[\Wb]: \tangentS[^2] \rightarrow \tangentS[^2] $  of tangential 2-tensor fields $ \qb[\para] $ in direction of $ \Wb\in\tangentR $ by
\begin{align}
    \gaugeJ[\Wb]\qb[\para] 
        &:= \gaugeS[\Wb]\qb[\para] - \Ab[\para,\Wb]\qb[\para] + \qb[\para]\Ab[\para,\Wb]
        \formPeriod
    \label{eq:jaumann_deformation_derivativ_tensor}
\end{align}
As with the previous definition \eqref{eq:jaumann_deformation_derivativ_vector} for vector fields, this also agrees with the 
corotational infinitesimal variation from \cite{SalbreuxJuelicherProstCallan-Jones__2022}.
The relation to the $ \natural $-convected deformation derivatives \eqref{eq:deformation_derivative_natural_tensor} is
\begin{align}
    \gaugeJ[\Wb]\qb[\para]
        &= \frac{1}{2}\left( \gaugeUU[\Wb]\qb[\para] + \gaugeLL[\Wb]\qb[\para] \right)
         = \frac{1}{2}\left( \gaugeUL[\Wb]\qb[\para] + \gaugeLU[\Wb]\qb[\para] \right)
    \formPeriod
    \label{eq:jaumann_deformation_derivativ_relation_to_convected_tensor}
\end{align}

\begin{definition}[Gauge of Jaumann Surface Independence on Tangential 2-Tensor Fields]\label{def:jauman_goi_tensor}
    A tangential 2-tensor field $ \qb[\para]\in \tangentS[^2] $ fulfill the gauge of Jaumann surface independence $ \gaugeJ\qb=0 $, if and only if
    one of the following equivalent statements is true:
    \begin{enumerate}[(i)]
    \item $ \gaugeJ[\Wb] \qb[\para] = 0 $ is valid for all $ \Wb\in\tangentR $.
    \item $ \gauge{\Wb}q^{ij}[\para] + g^{ik}[\para]g^{jl}[\para]\gauge{\Wb}q_{kl}[\para] = 0 $ is valid componentwise for the 
        fully contra- and covariant proxy field of $ \qb[\para] $ and for  all $ \Wb\in\tangentR $.
     \item $ g^{jk}[\para]\gauge{\Wb}\tensor{q}{^i_k}[\para] + g^{ik}[\para]\gauge{\Wb}\tensor{q}{_k^j}[\para] = 0 $ is valid componentwise for the 
        mixed contra- and covariant proxy field of $ \qb[\para] $ and for  all $ \Wb\in\tangentR $.
    \end{enumerate}
\end{definition}

As for tangential vector fields, restricting the deformation direction $ \Wb $ so that only rigid body deformations are considered,
\ie\ it is $ \Sb[\para,\Wb]=0 $, the Jaumann and all $ \natural $-convected deformation derivatives are equal.
In contrast, if we consider only strain deformations, \ie\ $ \Ab[\para,\Wb]=0 $, 
then the Jaumann and the material deformation derivatives are equal. 

\subsubsection{Consequences for Energy Variations}

All definitions, proceedings and consequences for energy variations in section \ref{sec:consequences_for_energy_variations_vector} are also applicable for
tangential 2-tensor fields, but for $ \qb[\para]\in\tangentS[^2] $.
Especially the chain rule \eqref{eq:deformation_transport_vector} reads
\begin{align}
    \inner[4]{\hilbertR}{\frac{\del\energy}{\del\para} , \Wb}
        &= \inner[4]{\hilbertR}{\frac{\partial\energy}{\partial\para} , \Wb}
            + \inner[4]{\hilbertS[^2]}{ \frac{\del\energy}{\del\qb},  \gauge[\surf]{\Wb}\qb[\para] }
     \label{eq:deformation_transport_tensor}
\end{align}
for an energy $ \energy = \energy[\para, \qb[\para]] $.
Moreover, identities \eqref{eq:deformation_change_vector} and \eqref{eq:deformation_change_vector_partial} become
\begin{align}
    \inner[4]{\hilbertR}{\frac{\del\energy}{\del\para} , \Wb} \hspace{-10pt}
        &= \inner[4]{\hilbertR}{\frac{\del\energy}{\del\para}\Big\vert_{\gauge[\Psi]{}\qb=0} , \Wb}
            \hspace{-15pt}+ \inner[4]{\hilbertS[^2]}{ \frac{\del\energy}{\del\qb},  \gauge[\Psi]{\Wb}\qb[\para] }
      \label{eq:deformation_change_tensor} \\
    \inner[4]{\hilbertR}{\frac{\partial\energy}{\partial\para} , \Wb} \hspace{-10pt}
        &= \inner[4]{\hilbertR}{\frac{\del\energy}{\del\para}\Big\vert_{\gauge[\Psi]{}\qb=0} , \Wb}
           \hspace{-15pt}+ \inner[4]{\hilbertS[^2]}{ \frac{\del\energy}{\del\qb},  \gauge[\Psi]{\Wb}\qb[\para] - \gauge[\surf]{\Wb}\qb[\para] },
      \label{eq:deformation_change_tensor_partial}
\end{align}
for symbols $ \Psi\in\{\surf,\sharp\sharp,\flat\flat,\sharp\flat,\flat\sharp,\jau\} $.

\subsubsection{Consequences for $ \hil $-Gradients}
The weak $ \hil $-gradient reads 
\begin{align*}
  &\innerH{\tangentR \times \tangentS[^2]}{ \hilgrad\energy, ( \Wb, \rb ) }\\
    &\quad\quad = \inner[4]{\hilbertR}{\frac{\partial\energy}{\partial\para} , \Wb}
          \hspace{-10pt}+ \inner[4]{\hilbertS[^2]}{\frac{\del\energy}{\del\qb} , \rb + \Psib_1[\para,\Wb]\qb[\para]+ \qb[\para]\Psib_2^T[\para,\Wb]}
       \formComma
\end{align*}
\begin{center}
\begin{tabular}{lr|c|c|c|c|c|cl}
 where &$ \Psib_1 = $ & $ 0 $ & $\Gb$ & $-\Gb^T$ & $\Gb$ & $-\Gb^T$ & $\Ab$   \\
 and   &$ \Psib_2 = $ & $ 0 $ & $\Gb$ & $-\Gb^T$ & $-\Gb^T$ & $\Gb$ & $\Ab$  \\
 for   &    $ 0 = $   & $ \gaugeS\qb $ & $ \gaugeUU\qb $ & $ \gaugeLL\qb $ & $ \gaugeUL\qb $ & $ \gaugeLU\qb $ & $ \gaugeJ\qb $&
\end{tabular}
\end{center}
for all $ \Wb\in\tangentR $ and $\rb\in\tangentS[^2]$.
Using $ \hil $-adjoints of $ \Psib_1[\para,\Wb],\Psib_2[\para,\Wb]\in\tangentS[^2] $
in compliance with \eqref{eq:deformation_adjoints}
we obtain the strong formulation
\begin{align}
    \hilgrad\energy
       &= \left( \frac{\del\energy}{\del\para}, \frac{\del\energy}{\del\qb}  \right)
       = \left( \frac{\partial\energy}{\partial\para} - \div_{\tangentS[^2]}\gstress, \frac{\del\energy}{\del\qb}  \right)
    \label{eq:L2Grad_strong_tensor}
\end{align}
with gauge stress $ \gstress\in\tangentS[^2] $ given by
\begin{align*}
    \gstress &=
      \begin{cases}
           0 & \text{for }  \gaugeS\qb = 0 \\
           \frac{\del\energy}{\del\qb}\qb^T[\para] +  \left(\frac{\del\energy}{\del\qb}\right)^T\qb[\para] =:\gstress_{\sharp\sharp}
                 & \text{for }  \gaugeUU\qb = 0 \\
           -\qb[\para] \left(\frac{\del\energy}{\del\qb}\right)^T - \qb^T[\para]\frac{\del\energy}{\del\qb} = -\gstress_{\sharp\sharp}^T
                & \text{for }  \gaugeLL\qb = 0 \\
           \frac{\del\energy}{\del\qb}\qb^T[\para]- \qb^T[\para]\frac{\del\energy}{\del\qb} =:\gstress_{\sharp\flat}
                & \text{for }  \gaugeUL\qb = 0 \\
           -\qb[\para] \left(\frac{\del\energy}{\del\qb}\right)^T +  \left(\frac{\del\energy}{\del\qb}\right)^T\qb[\para] = -\gstress_{\sharp\flat}^T
                & \text{for }  \gaugeLU\qb = 0 \\
           \frac{1}{2}\left( \gstress_{\sharp\sharp} -\gstress_{\sharp\sharp}^T \right) = \frac{1}{2}\left( \gstress_{\sharp\flat} -\gstress_{\sharp\flat}^T \right)
                & \text{for }  \gaugeJ\qb = 0 \formPeriod
      \end{cases} 
\end{align*}
Just as in the vector case, we point out that the gauge stress tensor depends on the syntactical description of the energy, since 
$ \frac{\partial\energy}{\partial\para} $ in \eqref{eq:L2Grad_strong_tensor} also depends on the syntax.
For instance, a semantically equal energy $ \tenergy[\para, q^{11}[\para], q^{12}[\para], q^{21}[\para], q^{22}[\para]] = \energy[\para, \qb[\para]]$,
with $ \qb[\para]=q^{ij}[\para]\partial_i\para\otimes\partial_j\para $,
gives a different gauge stress tensor $ \tgstress\neq\gstress $.  

\subsubsection{Consequences for $ \hil $-Gradient Flows}
The associated $ \hil $-gradient flow $\left( \Vb[\para], \timeD\qb[\para] \right) = -\lambda \hilgrad\energy$
can be formulated with various observer-invariant instantaneous time derivative $ \timeD:\tangentS[^2]\rightarrow\tangentS[^2] $ given in \cite[Conclusion 7]{Nitschke_2022}. We treat in this section the material ($ \dot{\qb} $), upper-upper or fully-upper ($ \timeLuu\qb $), lower-lower or fully-lower ($ \timeLll\qb $),  
upper-lower ($ \timeLul\qb $), lower-upper ($ \timeLlu\qb $)  and Jaumann ($ \timeJ\qb $)
time derivative of a tangential 2-tensor field $ \qb$.
For a material (Lagrange) observer, the material time derivative reads
\begin{align*}
    \dot{\qb} 
        &= (\partial_t q^{ij}[\para])\partial_i\para\otimes\partial_j\para
            + \Gb[\para,\Vb[\para]]\qb[\para] + \qb[\para]\Gb^T[\para,\Vb[\para]]\\
        &=\proj_{\tangentS[^2]} \frac{d}{dt}\qb[\para] \formPeriod
\end{align*}
All other time derivatives can be derived from this one by
\begin{align*}
    \timeD\qb[\para] &= \dot{\qb}[\para] - \Phib_1[\para,\Vb[\para]]\qb[\para] - \qb[\para]\Phib_2^T[\para,\Vb[\para]]
    \formComma
\end{align*}
\begin{center}
\begin{tabular}{lr|c|c|c|c|c|cl}
 where &$ \Phib_1 = $ & $ 0 $ & $\Gb$ & $-\Gb^T$ & $\Gb$ & $-\Gb^T$ & $\Ab$   \\
 and   &$ \Phib_2 = $ & $ 0 $ & $\Gb$ & $-\Gb^T$ & $-\Gb^T$ & $\Gb$ & $\Ab$  \\
 for   &    $ \timeD\qb = $   & $ \dot{\qb} $ & $ \timeLuu\qb $ & $ \timeLll\qb $ & $ \timeLul\qb $ & $ \timeLlu\qb $ & $ \timeJ\qb $&,
\end{tabular}
\end{center}
see \cite{Nitschke_2022}.
The time-depending solution $ (\para, \qb[\para]) $ of the gradient flow yields the energy rate
 \begin{align}
    \frac{d}{dt}\energy
        &= - \lambda\left\| \hilgrad\energy \right\|_{\hil(\tangentR\times\tangentS[^2])}^2\\
        &\quad + \innerH[2]{\tangentS[^2]}{\frac{\del\energy}{\del\qb} , (\Phib_1-\Psib_1)[\para,\Vb[\para]]\qb[\para] + \qb[\para] (\Phib_2-\Psib_2)^T[\para,\Vb[\para]]}
    \formPeriod
    \label{eq:dissipation_tensor}
\end{align}
Therefore, a consistent choice of gauge and time derivative, \ie\ $ (\Psib_1,\Psib_2) = (\Phib_1,\Phib_2) $ 
would ensure a decreasing energy in the $ \hil $-gradient flow.
This choice is also recommended, if we want the $ \hil $-gradient flow to be consistent with 
Onsagers variational principle \cite{Doi}, \cf\ remark \ref{rem:onsager_vector}.

In section \ref{sec:example_tensor} we briefly mention an example for a Landau-de Gennes-Helfrich energy. 

\section{Examples}
\label{sec:Examples}

\subsection{Examples for scalar fields}\label{sec:example_scalar}

We here select several examples for scalar fields on deformable surfaces. They range from particle density fields, \eg\ protein interactions in viral capsides \cite{Aland_MMS_2012}, colloids at fluid-fluid interfaces in emulsions and bijels \cite{Aland_PF_2011,Aland_PRE_2012} and adatoms on material surfaces \cite{Burger_CMS_2006, Raetz_Nonl_2007}, to lipid and protein concentrations in biomembranes \cite{Wang_JMB_2008,Lowengrub_PRE_2009,Elliott_JCP_2010,Elliott_ARMA_2016} and models for cell migration \cite{Jilkine_PLOSCB_2011,Marth_JMB_2014}, to applications on larger length scales in materials science, \eg\ dealloying by surface dissolution \cite{Eilks_JCP_2008} or grain boundary diffusion \cite{Hoetzer_AM_2019}. The list can be extended to other contributions for these examples and further applications.

\subsection{Minimizing a Frank-Oseen-Helfrich Energy}\label{sec:example_vector}

    We consider a one-constant approximation of the surface Frank-Oseen(FO) energy, $\energy_{\nabla}[\para,\qb[\para]] + \energy_{R}[\para,\qb[\para]]$ \cite{Nitschke_2019}, together with the Helfrich energy with vanishing spontaneous curvature, $\energy_{H}[\para]$, \cite{Helfrich_1973}. The energies read:
    \begin{align*}
        \energy[\para,\qb[\para]] 
            &:= \energy_{\nabla}[\para,\qb[\para]] + \energy_{R}[\para,\qb[\para]] + \energy_{H}[\para]\\
        \energy_{\nabla}[\para,\qb[\para]]
            &:= \frac{K}{2}\left\| \nabla\qb[\para] \right\|^2_{\hilbertS[^2]} \\
        \energy_{R}[\para,\qb[\para]]    
            &:= \frac{K}{2}\left\| \shop[\para]\qb[\para] \right\|^2_{\hilbertS}
                +  \frac{\omega}{4}\left\| \left\| \qb[\para] \right\|^2_{\tangentS} - 1 \right\|^2_{\hilbertS[^0]}\\
        \energy_{H}[\para]
            &:= \frac{\kappa}{2} \left\| \meanc[\para] \right\|^2_{\hilbertS[^0]}
    \end{align*}
    with $ K,\omega,\kappa > 0 $.
    The FO energy forces a solution $ \qb[\para]\in\tangentS $ to be parallel, aligned along lines of minimal curvature, and to be normalized.
    For $ \omega\rightarrow\infty $ such a solution can be interpreted as a (polar) director field almost everywhere.
    Due to the Helfrich energy, the surface attempts to minimize its mean curvature. For practical applications one might need to add constraints on area conservation, see \cite{Reuteretal_PRSA_2019_Preprint}.
    The unusual splitting of the FO energy into the $ \nabla $-part and $ R $(emainder)-part is due to the following calculation, in which we treat
    both parts differently.
    Equivalently, one could write the elastic contributions, the two terms with paramter $ K $ together as an extrinsic distortion energy 
    $ \frac{K}{2}\left\| \nabla_\surf\qb[\para] \right\|^2_{\hilbert{\tangentR\otimes\tangentS}} $ containing the surface derivative \eqref{eq:surface_derivative}, see \cite{Morris_2022}. Other related formulations are considered in \cite{Kralj_2011,Lopez-Leon_2011,Napoli_2012,Segatti_2016,Nestler_2018}. However, they all don't consider $\para$ as a degree of freedom.
    
    The variations \wrt\ $ \qb[\para] $ can be obtained in the usual way and read
    \begin{align*}
        \frac{\del\energy_{\nabla}}{\del\qb} 
            &= -K\Delta\qb[\para] \\
        \frac{\del\energy_{R}}{\del\qb} 
            &=  K\shop^2[\para]\qb[\para] + \omega\left( \left\|\qb[\para] \right\|^2_{\tangentS} - 1 \right)\qb[\para] 
    \end{align*}
    in their strong formulations, where $ \Delta := -\nabla^{\adj}\circ\nabla = \div\circ\nabla $ is the Bochner-Laplace operator.
    The Helfrich energy does not depend on $ \qb[\para] $, and thus, $\frac{\del\energy_{H}}{\del\qb} = \nullb$.
    
    Note that the overall energy has the form 
    $ \energy[\para,\qb[\para]] = \int_{\surf} u[\para,\qb[\para]] \mu[\para] $ with an energy density 
    $ u[\para,\qb[\para]]\in\tangentS[^0] $ and due to this yields
    \begin{align}
        \inner[4]{\hilbertR}{\frac{\del\energy}{\del\para} , \Wb} \hspace{-10pt}
            &= \int_{\surf} \gauge{\Wb}u[\para,\qb[\para]]  + u[\para,\qb[\para]]\Tr\Gb[\para,\Wb] \mu[\para]
        \label{eq:deformation_transport_theorem}
    \end{align}
    as a consequence of $ \gauge{\Wb}\mu[\para] $ given in appendix \ref{sec:deformation_of_metric_density}.
    This allows us to use deformation derivatives to determine variations \wrt\ $ \para $ for such kind of energies.
    Moreover, we take advantage of \eqref{eq:deformation_change_vector}, \resp\ \eqref{eq:deformation_change_vector_partial}, 
    which allows to variate the $ \nabla $- and $ R $-part
    under certain gauges of surface independence and obtain a general result.
    Note that the material deformation derivative $ \gaugeS[\Wb] $ is metric, \resp\ inner product, compatible due to \eqref{eq:deformation_derivativ_tensor} and 
    \eqref{eq:deformation_of_inverse_metric}.
    Under the assumption of $ \gaugeS \qb = 0 $ and with \eqref{eq:deformation_of_shape_operator} the $ R $-part yields
    \begin{align*}
        \inner[4]{\hilbertR}{\frac{\del\energy_{R}}{\del\para}\Big\vert_{\gaugeS\qb=0} , \Wb} \hspace{-80pt}\\
            &= K\inner{\hilbertS[^2]}{(\gaugeS[\Wb]\shop[\para])\qb[\para],\shop[\para]\qb[\para]}\\
            &\quad + \inner{\hilbertS[^0]}{u_{R}[\para,\qb[\para]], \Tr\Gb[\para,\Wb]}\\
            &= K\inner{\hilbertS[^2]}{\shop[\para]\qb[\para]\otimes\qb[\para], \nabla \bb[\para,\Wb]}\\
            &\quad -\inner{\hilbertS[^2]}{K\shop^2[\para]\qb[\para]\otimes\qb[\para] - u_{R}[\para,\qb[\para]]\Id_{\tangentS}, \Gb[\para,\Wb]}  \formPeriod
    \end{align*} 
    where $ \Gb[\para,\Wb]\in\tangentS[^2] $ and $ \bb[\para,\Wb]\in\tangentS $ are the effective tangential- and non-tangential-parts of the surface gradient
    $ \nabla_{\surf}\Wb\in\tangentR[^2] $ given in \eqref{eq:surface_derivative_of_W}.
    Eventually by  \eqref{eq:deformation_change_vector_partial}, we can state the gauge independent partial variation
    \begin{align}
        &&\inner[4]{\hilbertR}{\frac{\partial\energy_R}{\partial\para} , \Wb} \hspace{-15pt}
            &= -\inner{\hilbertR[^2]}{\sigmab_{R} + \normal[\para]\otimes\div\sigmab_{R,K} , \nabla_{\surf}\Wb}\\
        &\text{with}&\sigmab_{R}
            &:= \shop[\para]\sigmab_{R,K} - u_R[\para,\qb[\para]]\Id_{\tangentS} \in\tangentS[^2]\notag\\
        &\text{and}&\sigmab_{R,K}
            &:= K \shop[\para]\qb[\para]\otimes\qb[\para]\in\tangentS[^2] \notag \formPeriod
    \end{align}
    
    The difficulty in the $ \nabla $-part is that spatial and deformation derivatives are not commuting in general.
    But for $ \gaugeU\qb=0 $, \resp\ $\gauge{} q^{i}=0$, we can use that $ \gauge{\Wb}\partial_k q^{i}[\para] = 0 $ as a consequence.
    With a detailed calculation in appendix \ref{sec:calculation_for_example},
    identity \eqref{eq:deformation_change_vector_partial} and the upper-convected deformation derivative \eqref{eq:upper_deformation_derivativ_vector},
    we obtain the gauge independent partial variation
    \begin{align}
        \inner[4]{\hilbertR}{\frac{\partial\energy_\nabla}{\partial\para} , \Wb} \hspace{-10pt}
            &=  \inner[4]{\hilbertR}{\frac{\del\energy_{\nabla}}{\del\para}\Big\vert_{\gaugeU\qb=0} , \Wb} 
                \hspace{-15pt}- \inner[4]{\hilbertS[^2]}{ \frac{\del\energy_{\nabla}}{\del\qb}\otimes\qb[\para],  \Gb[\para,\Wb] }\notag\\
            &= -\inner{\hilbertR[^2]}{ \sigmab_{E,I} 
                    + \normal[\para]\otimes\etab_{\nabla}, \nabla_{\surf}\Wb }
     \label{eq:partial_variation_UNabla}\\
        \text{where }\quad \sigmab_{E,I}
            &= K \left( (\nabla\qb[\para])^T \nabla\qb[\para] - \frac{\left\| \nabla\qb[\para] \right\|^2_{\tangentS[^2]}}{2} \Id_{\tangentS} \right)\notag\\
        \text{and }\quad\quad \etab_{\nabla} 
            &= K \left( (\nabla_{\shop[\para]}\qb[\para])\qb[\para] - \nabla_{(\shop[\para]\qb[\para])}\qb[\para]\right) \notag\formComma
    \end{align}
    where $ \sigmab_{E,I}\in\Qcal^2\surf $ is the intrinsic, trace-free and symmetric Ericksen tensor.
    
    As a consequence of the local mean curvature deformation \eqref{eq:deformation_of_mean_curvature} and identity \eqref{eq:deformation_transport_theorem}, the variation of the Helfrich energy leads to
    \begin{align*}
        \inner[5]{\hilbertR}{\frac{\partial\energy_H}{\partial\para} , \Wb}  \hspace{-20pt}
            &= -\kappa\left( \inner[5]{\hilbertS}{\nabla\meanc[\para],\bb[\para,\Wb]} 
                        +  \inner[5]{\hilbertS[^2]}{ \meanc[\para]\left( \shop[\para] - \frac{\meanc[\para]}{2}\Id_{\tangentS} \right), \Gb[\para,\Wb] }\right)\\
            &= \kappa \inner[4]{\hilbertR}{ \left( \Delta\meanc[\para] + \frac{\meanc[\para]}{2} \left( \meanc^2[\para] - 4\gaussc[\para] \right) \right) \normal[\para] , \Wb} \formComma
    \end{align*}
    with the inner tangential component completely canceling out, since the shape-operator is curl-free, \ie\ $\div\shop[\para] = \nabla\meanc[\para] $.
    It is not particularly surprising that a purely geometric measure induces forces exclusively in the normals direction.
    
    Finally, with surface divergence \eqref{eq:surface_divergence}, the  $ \hil $-gradient flow \eqref{eq:L2flow} containing the $ \hil $-gradient \eqref{eq:L2Grad_strong} yields
    \begin{align}
        \Vb[\para] &= -\div_{\surf}\left( \sigmab_{\!FO} - \gstress + \normal[\para]\otimes\etab_{FO} \right) + f_H\normal[\para],
        \quad\quad\timeD\qb[\para] = -\frac{\del\energy}{\del\qb}\formComma
            \label{eq:FOModel}\\
        \frac{\del\energy}{\del\qb} &= -K(\Delta\qb[\para] - \shop^2[\para]\qb[\para]) + \omega\left( \left\|\qb[\para] \right\|^2_{\tangentS} - 1 \right)\qb[\para] \formComma
        \label{eq:FOMolecular}
    \end{align}
    with
    \begin{align}
        \sigmab_{\!FO}
            &= K\left(  (\nabla\qb[\para])^T \nabla\qb[\para] + \shop^2[\para]\qb[\para]\otimes\qb[\para]\right) - u\Id_{\tangentS}\formComma\notag \\
        u 
            &= \frac{K}{2}\left(\left\| \nabla\qb[\para] \right\|^2_{\hilbertS[^2]}
                            + \left\| \shop[\para]\qb[\para] \right\|^2_{\tangentS}\right)
                            +  \frac{\omega}{4} \left(\left\| \qb[\para] \right\|^2_{\tangentS} - 1 \right)^2\formComma\notag\\
        \etab_{FO} 
            &= K \left((\nabla_{\shop[\para]}\qb[\para])\qb[\para]  - \nabla_{(\shop[\para]\qb[\para])}\qb[\para]
                    + \div( \shop[\para]\qb[\para]\otimes\qb[\para] )\right) \notag \formComma \\
        f_H
            &= -\kappa \left( \Delta\meanc[\para] + \frac{\meanc[\para]}{2} \left( \meanc^2[\para] - 4\gaussc[\para] \right) \right)
    \end{align}
    a chosen time derivative $ \timeD:\tangentS\rightarrow\tangentS $ \eqref{eq:time_derivatives_vector}
    and a gauge stress tensor field $ \sigmab\in\tangentS[^2] $ according to table \ref{tab:gauge_stress} depending on the chosen gauge of surface independence.
    Note that we set the mobility parameter $ \lambda=1 $.
    
    For analytical reasons, we could also apply a tangential-normal-splitting on $ \frac{\del\energy}{\del\para} $.
    With the aid of the splitting \eqref{eq:surface_divergence} of the surface divergence
    and orthogonal decomposition $ \Vb[\para] = \vb[\para]+\vnor[\para]\normal[\para]\in\tangentS\oplus(\tangentS[^0])\normal[\para] $, 
    we can represent the spatial evolution equation in \eqref{eq:FOModel} by
    \begin{align}
        \vb[\para]
            &= \div \gstress + \frac{\del\energy}{\del\qb}\nabla\qb[\para] \formComma
         \label{eq:FOModel_tangential}\\
        \vnor[\para]
            &= \inner{\tangentS[^2]}{ \gstress -\sigmab_{E}, \shop[\para]} 
                        - \div\etab_{FO} + \frac{\omega}{4}\meanc\left(\left\| \qb[\para] \right\|^2_{\tangentS} - 1 \right)^2
                        + f_H \formComma
         \label{eq:FOModel_normal}\\
        \sigmab_{E}
            &= K \proj_{\mathcal{Q}^2\surf}\left( (\nabla\qb[\para])^T\nabla\qb[\para] + \shop[\para]\qb[\para] \otimes  \shop[\para]\qb[\para] \right)\notag
    \end{align}
    as a 
    consequence
    of Weitzenböck-identity, Gauss-identity and curl-freeness of the second fundamental form.
    The tensor $ \sigmab_{E}\in\mathcal{Q}^2\surf < \tangentS[^2] $ is called the trace-free surface Ericksen-stress \cite{Reuteretal_PRSA_2019_Preprint}.
    This tensor occurs similarly in flat hydrodynamic liquid crystal models, \eg\ in \cite{Lin_1995}.
    To make the comparison easier we could use the identity
    $ \frac{\sigmab_{E}}{K} + p_{E}\Id_{\tangentS} = (\nabla_{\surf}\qb[\para])^T\nabla_{\surf}\qb[\para] $,
    where $ p_{E}= \frac{\|\nabla_{\surf}\qb[\para]\|^2}{2} $ can be included into a generalized pressure term.
    
    In \cite{Reuteretal_PRSA_2019_Preprint} (a preprint of \cite{Reuteretal_PRSA_2020})
    a similar $\hil$-gradient flow is given for the material gauge of surface independence and material time derivative, but restricted to evolutions in normal direction.
    The calculated $\hil$-gradient flow and energy rate equals our equations \eqref{eq:FOModel}(right)+\eqref{eq:FOModel_normal} and energy rate  \eqref{eq:dissipation} for the given restriction, neglecting the surface area penalization part.
    It should be noted for the comparison that 
    $ \div\etab_{FO} = K\div((\nabla_{\qb[\para]}\meanc[\para]+2\nabla_{\shop[\para]}\qb[\para])\qb[\para]) $ holds, 
    according to \cite[Proposition B.1.]{Reuteretal_PRSA_2019_Preprint}.
    
\begin{figure}[t]
        \centering
        \includegraphics[width=0.99\linewidth]{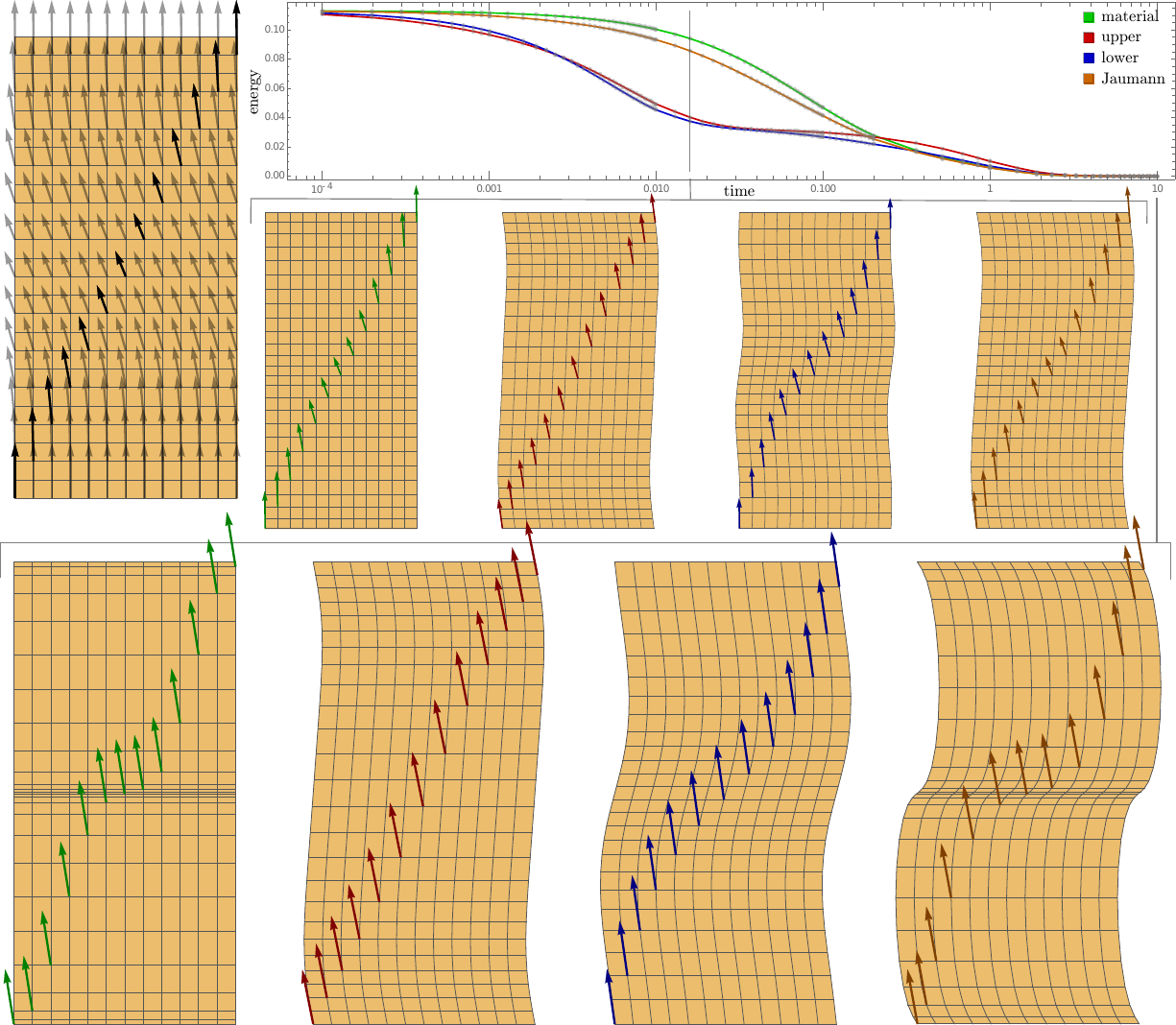}       
        \caption{Solutions for the $ \hil $-gradient flows \eqref{eq:FO_flat} with $ K = 0.1 $ and $ \omega=0.5 $: 
                 The initial solution \emph{(top left)} is periodic on the boundaries and constant horizontally.
                 Vertically, it is twisted counterclockwise and compressed towards the center.
                 For a better readability, vector field solutions are only plotted on a diagonal without loss of information.
                 The solutions are approximated for four different models, where we assume the gauge of surface independence and time derivative consistently,
                 namely the {\color{green} material},  {\color{red} upper-convected}, {\color{blue} lower-convected} and {\color{orange} Jaumann} gauge/derivative
                 \emph{(from left to right)}.
                 All energy paths \emph{(top right)} are decreasing as expected by energy rate \eqref{eq:dissipation} with $ \Phib=\Psib $.
                 The energy plot has a log-scale on its time axis.
                 Vector field solutions tend to be parallel aligned with a constant length towards the time process.
                 We show the solutions at time $ t = 0.016 $ \emph{(middle row)} and $ t = 10 $ \emph{(bottom row)}, where they almost reached their minimal state.
                 The time-step size $ \tau $ of the time discretization increases from $ \tau = 10^{-4} $ at the beginning until
                 $ \tau = 0.45 $ at the end, see the semitransparent gray dots at the energy paths.
                 The grid size $ h = 10^{-2} $ is constant along the vertical local Lagrange coordinate $ y^2 $ for all times.
                 The horizontal gray gridlines only represent every fourth available spatial line, 
                 \ie\ the distance between two gridlines is $ 0.04 $ constantly on local Lagrange coordinates.
                 Solutions are plotted \wrt\ Euler coordinates, \ie\ \wrt\ the embedding space, to show the local deformations of the underlying domain towards the minimum state.
                 Any impressions that the surface gets bent in normal direction is just an optical illusion.
                 In absence of normal forces the surface stays flat all times.
                 (Video online: \cite{mendeley_video})}
        \label{fig:experiment}
        \end{figure}

    The fact that the choice of the gauge of surface independence and the observer-invariant time derivative for the $ \hil $-gradient flow 
    $  \left( \Vb[\para], \timeD\qb[\para] \right) = -\lambda \hilgrad\energy $ \eqref{eq:L2flow} has an impact on its solution, can be seen from the energy rate \eqref{eq:dissipation} alone.
    In order to demonstrate the differences we simplify the problem. Certainly the curvature of a surface is a large factor, but the fundamental connection between the dynamic of a gradient flow system and the gauge of surface independence or time derivative is not curvature driven. 
    We therefore use an initial condition, which ensures flatness of the surface, namely $ \shop[\para]\vert_{t=0} = \nullb $ 
        and $ \vnor[\para]\vert_{t=0} = \inner{\tangentR}{\Vb[\para],\normal[\para]}\vert_{t=0} = 0 $.
        These flatness-symmetry conditions are preserved in time, since the only force in normal direction is the vanishing part \eqref{eq:FOModel_normal}. As a conserquence the Helfrich energy no longer contributes.
        Therefore the gradient flow only comprises the tangential equation 
        \eqref{eq:FOModel_tangential}.
        This yields the reduced equations of motion
        \begin{align}
            \vb[\para] 
                &=  \div \gstress + \frac{\del\energy}{\del\qb}\nabla\qb[\para] \formComma\quad\quad
            \timeD\qb[\para] 
                = -\frac{\del\energy}{\del\qb} \formComma
            \label{eq:FO_flat}\\
            \frac{\del\energy}{\del\qb}
                &=  -K\Delta\qb[\para] + \omega\left( \left\|\qb[\para] \right\|^2_{\tangentS} - 1 \right)\qb[\para] \notag
        \end{align}
        with a choice of time derivatives \eqref{eq:time_derivatives_vector} and gauge stress tensors $ \gstress\in\tangentS[^2] $ given in table \ref{tab:gauge_stress}.
        The initial surface is a rectangle with opposing periodic boundaries, see figure \ref{fig:experiment} (top left), \ie\ it can be seen as a flat torus or an infinite plane with a periodic pattern.
        To keep track of local deformations, we use the material parameterization approach
        $ \para(t,y^1,y^2) = [ y^1 + f^1(t,y^1,y^2), y^2 + f^2(t,y^1,y^2), 0 ]^T =[x,y,z]^T$ with $ f^1(0,y^1,y^2)=f^2(0,y^1,y^2)=0 $,
        \ie\ the observer is taken the Lagrange-perspective and $ f^1 $, \resp\ $ f^2 $, describe the local deformation in $ x $-, \resp\ $ y $-direction, of the initial rectangle.
        We assume that the initial condition for $ \qb[\para]\vert_{t=0}=(q^i[\para]\partial_i\para)\vert_{t=0}\in\tangentS\vert_{t=0} $ bears the same periodicity as the surface.
        Moreover, we stipulate that $\qb[\para]\vert_{t=0}$ is constant along $ y^1 $-coordinate lines.
        Since also the parameterization sustains the same symmetry, the model \eqref{eq:FO_flat} preserve the independency of $ y^1 $ in time.
        This has the advantage, that a spatial discretization has to be done along $ y^2 $ only.
        Eventually, our degrees of freedom are $ \{f^1(t,y^2), f^2(t,y^2), q^1[\para](t,y^2), q^2[\para](t,y^2)\} $ 
        for all $ y^2 $ and $ t $ in their considered domains under the discussed initial conditions above.
        The minimal state of $ \energy $ comprises a spatially parallel vector field with constant length.
        Since the periodic surface is topologically equivalent to a torus, \ie\ the Euler characteristic is $ \chi(\surf) = 0 $,
        this minimal states do not contain any defects.
        Therefore, a stationary minimal solution depends neither on the gauge of surface independence nor on a time derivative.
        But out of this state of equilibrium, the force $ \frac{\del\energy}{\del\para} $ clearly depends on the gauge 
        as well as the dynamic counterpart of $ \frac{\del\energy}{\del\qb} $ depends the choice of time derivative.
        The big impact of these choices on the evolution of \eqref{eq:FO_flat} can be seen in figure \ref{fig:experiment}.
        For consistent choices, all models evolve to a minimal state, where $\energy$ is vanishing, but the paths are differently and so the final surface.  
        
        The highly nonlinear equations \eqref{eq:FO_flat} is discretized by an implicit Euler scheme in time, 
        where we linearize all terms by one step Taylor expansion at the old time step. 
        We use FDM with central differential quotients for any order of spatial derivatives.
        Solving the resulting discrete linear system is done with the \texttt{UMFPACK} library in every time step.
        Unfortunately, it was only possible to get stable solutions for a consistent choice of gauge and time derivative with this proceeding.
        Therefore, the energy rate \eqref{eq:dissipation} becomes 
        $ \frac{d}{dt}\energy = - \lambda\left\| \hilgrad\energy \right\|_{\hil(\tangentR\times\tangentS)}^2 \le 0 $, 
        \ie\ the analytical solutions guarantee dissipation.
        This is also reflected by the numerical solutions, see figure \ref{fig:experiment}.

\subsection{Minimizing Landau-de Gennes-Helfrich Energy}\label{sec:example_tensor}

A $ \hil $-gradient flow is given in \cite{Reuteretal_PRSA_2020} for the Landau-de Gennes-Helfrich energy, where $ \qb[\para]\in\mathcal{Q}^2\surf < \tangentS[^2] $ is
a trace-free and symmetric 2-tensor field and surface deformations are restricted to deformations in normal direction. The corresponding surface energy $ \energy[\para,\qb[\para]] $ was derived as a thin film limit of a three-dimensional energy \cite{Nitschkeetal_PRSA_2018}. The resulting equations of motion are formulated under the material gauge of surface independence and with the material time derivative.
The calculated energy rate also fits the energy rate \eqref{eq:dissipation_tensor} for the given restrictions.

\section{Conclusions}
\label{sec:Conclusions}

Tangential tensor fields are usually defined on their underlying surface. However, if the surface smoothly changes, the field also has to be defined on the deformed surface, e.g., through various pushforwards. The linear change of these definitions is quantified by the so-called deformation derivatives and can be seen as local domain variations of the first kind of the fields. Similarly to time derivatives, deformation derivatives can be defined in different ways. We presented one for scalar fields, four for tangential vector fields, and six for tangential 2-tensor fields. With these deformation derivatives, we can quantify what it means for a tensor field to be independent of the underlying surface, which leads to different definitions of the gauges of surface independence, where deformation derivatives vanish for all possible deformations of the surface. For non-scalar fields, further possibilities exist to define deformation derivatives and thus gauges of surface independence. However, those chosen already cover a wide range of physically meaningful applications. Moreover, we disregard handling $ n $-tensor fields for $ n>2 $ for reasons of space. It does not pose any difficulties to do it for fixed $ n $ in the same way as for $n = 1,2$, although it becomes quite technical for general tensor orders. Potential applications are hexatic orders ($n = 6$) in confluent cell layers \cite{armengol2023epithelia} or quad mesh generation ($n = 4$) in computer graphics \cite{Bommes2013}. Also, an extension to $ d $-dimensional hypersurfaces smoothly embedded in a higher-dimensional Euclidean space, with co-dimension greater than one, should not pose a major challenge.

We see the primary application for the gauges of surface independence in the derivation of $\hil$-gradient flows of energies, where the phase space comprises the surface itself among tensor quantities defined on it.
In order to guarantee a decreasing energy in the $ \hil $-gradient flow, the time derivative and the gauge of surface independence have to be chosen consistently. The choice should thereby follow physical arguments. We suggest imagining how the considered tensor field would act under force-free transport and selecting a time derivative and a gauge of surface independence that fit the expected physical behavior, see, e.g., \cite{Nitschke_2022}. Another possibility is to contemplate a physically suitable pushforward for the tensor field, as its linear information determines a gauge of surface independence. Both provide a systematic way to select an appropriate $ \hil $-gradient flow.

Note that equilibrium states, stable or not, are independent of the choice of the gauge of surface independence, see the discussion above \eqref{eq:L2flow}. But in many situations, these states are not unique. Therefore, steepest descent methods, like the $ \hil $-gradient flow, which approach the equilibrium state from non-equilibrium states, can give quite different results for different gauges of surface independence. This dependence on the gauge of surface independence has been shown by an example in section \ref{sec:example_vector}.

Besides these physical implications, this work also provides some helpful tools for variations of the first kind. Identity \eqref{eq:deformation_change_vector}, and its special case \eqref{eq:deformation_transport_vector}, allow for variation under a specific gauge of surface independence, even though the tensor field actually obeys a different gauge. We used this for the example in section \ref{sec:example_vector} to circumvent calculating all commutators with respect to the covariant spatial and deformation derivative. Instead, we could derive the variation only for a single gauge, which seems to simplify the situation, and concluded on the general case afterward. We'd like to mention that the total variation $ \frac{\partial \energy}{\partial \para} $, deformation derivatives, and hence also the gauges, follow a general principle of covariance, i.e., they are independent of their syntactical description, e.g., partial dependencies in $ \energy $ or coordinate dependencies. Contrarily, the strong formulation of the $ \hil $-gradient \eqref{eq:L2Grad_strong}, in terms of partial variation and gauge stress, is certainly helpful, but both summands have to be interpreted in the context of the syntax of $ \energy $. We showed in Appendices \ref{sec:upper_deformation_transport_vector} and \ref{sec:deformation_transport_vector_with_metric} that this has no influence on the left-hand side of \eqref{eq:L2Grad_strong}.



\appendix

\section{Alternative Proceedings}

\subsection{Contravariant Proxy Approach to \eqref{eq:deformation_transport_vector}}\label{sec:upper_deformation_transport_vector}
Instead of considering tangential vector fields $ \qb[\para]\in\tangentS $ together with an associated frame, some readers may prefer to
use the contravariant proxy field $ \{q^i[\para]\}\in(\tangentS[^0])^2$,
where in case of an embedding with parameterization $ \para $ holds $ q^i[\para]\partial_i\para = \qb[\para] $,
\resp\ $ q^i[\para] = g^{ij}[\para]\inner{\tangentS}{\qb[\para] , \partial_j\para} $.
Therefore, we can rewrite the energy $ \energy $  in section \ref{sec:consequences_for_energy_variations_vector} by
\begin{align}
    \tenergy[\para,q^1[\para],q^2[\para]] 
        &:= \energy[\para, \qb[\para]]
    \label{eq:energy_subst_cov}
\end{align}
without changing the sematic of $ \U $.
The great advantage of total derivations, such as the total variation \eqref{eq:variation_X_raw_vector}, is that they are syntactical invariant, 
\ie\ the total variation
\begin{align*}
    \left\langle \frac{\del\tenergy}{\del\para},  \Wb \right\rangle
            &:= \lim_{\eps\rightarrow 0 } \frac{\tenergy[\para + \eps\Wb, q^1[\para + \eps\Wb],  q^2[\para + \eps\Wb]] - \tenergy[\para, q^1[\para],q^2[\para]]}{\eps}
\end{align*}
in an arbitrary direction $ \Wb\in\tangentR $ equals $ \left\langle \frac{\del\energy}{\del\para},  \Wb \right\rangle $ \eqref{eq:variation_X_raw_vector} as we can 
easily see by substitution \eqref{eq:energy_subst_cov}.
The total variations \wrt\ contravariant proxy components in direction of arbitrary scalar fields $ f\in\tangentS[^0] $ are
\begin{align*}
    \left\langle \frac{\del\tenergy}{\del q^1},  f \right\rangle
                &:= \lim_{\eps\rightarrow 0 } \frac{\tenergy[\para, q^1[\para] + \eps f, q^2[\para]] - \tenergy[\para,q^1[\para],q^2[\para]]}{\eps}\formComma\\
    \left\langle \frac{\del\tenergy}{\del q^2},  f \right\rangle
                    &:= \lim_{\eps\rightarrow 0 } \frac{\tenergy[\para, q^1[\para], q^2[\para] + \eps f] - \tenergy[\para,q^1[\para],q^2[\para]]}{\eps} \formPeriod
\end{align*}
Since $ \energy[\para,\qb[\para]+\eps\rb] = \tenergy[\para,q^1[\para] + \eps r^1,q^2[\para] + \eps r^2] $ is valid for 
an arbitrary tangential field $ \rb=r^i\partial_i\para\in\tangentS $, 
the relation to the total variation \eqref{eq:variation_q_raw_vector} is
\begin{align*}
    \left\langle \frac{\del\energy}{\del\qb},  \rb \right\rangle
        &= \left\langle \frac{\del\tenergy}{\del q^1},  r^1 \right\rangle + \left\langle \frac{\del\tenergy}{\del q^2},  r^2 \right\rangle 
         =  \left\langle \frac{\del\tenergy}{\del q^i},  r^i \right\rangle\formPeriod
\end{align*}
To get a decomposition as \eqref{eq:deformation_transport_vector} to separate determined and not determined parts, 
the proceeding is almost the same.
We use a Taylor expansion with a reasonable pushforward for scalar fields, \ie\ it holds
\begin{align}
    q^i[\para + \eps\Wb] 
        &= (q^i[\para])^{*_{\eps\Wb}} + \eps(\gauge{\Wb}q^i[\para])^{*_{\eps\Wb}} + \landau(\eps^2)\notag\\
    \forall f\in\tangentS[^0]:\quad f^{*_{\eps\Wb}}(y^1,y^2)
        &:= f(y^1,y^2) \in \tangentStar[^0_{(\para+\eps\Wb)(y^1,y^2)}]{\surf_{\eps\Wb}}
    \label{eq:pushforward_scalar}
\end{align}
with scalar-valued deformation derivative $ \gauge{\Wb}: \tangentS[^0] \rightarrow \tangentS[^0] $ \eqref{eq:deformation_derivativ_scalar}.
Another Taylor expansion and exploiting the vector space structure of an consistently chosen Hilbert space gives
\begin{align*}
    \tenergy[\para + \eps\Wb, q^1[\para + \eps\Wb],  q^2[\para + \eps\Wb]] \hspace{-100pt}\\
        &= \tenergy[\para + \eps\Wb, (q^1[\para])^{*_{\eps\Wb}},  (q^2[\para])^{*_{\eps\Wb}}]\\
        &\quad +\eps  \inner[4]{\hilbert{\tangent[^0]{\surf_{\eps\Wb}}}}
                                {\frac{\del\tenergy}{\del (q^i)^{*_{\eps\Wb}}}, (\gauge{\Wb}q^i[\para])^{*_{\eps\Wb}}}
          +\landau(\eps^2)\formPeriod
\end{align*}
Since $ *_{\eps\Wb} \rightarrow \Id_{\tangentS[^0]} $ is valid for $ \eps \rightarrow 0 $, 
we obtain the total variation
\begin{align}
    \innerH[4]{\tangentR}{ \frac{\del\tenergy}{\del\para},  \Wb}
        &= \innerH[4]{\tangentR}{ \frac{\partial\tenergy}{\partial\para},  \Wb}
           + \innerH[4]{\tangentS[^0]}{\frac{\del\tenergy}{\del q^i}, \gauge{\Wb}q^i[\para] }
     \label{eq:deformation_transport_vector_contarvariantly}
\end{align}
by defining the partial variation
\begin{align*}
   \inner{}{ \frac{\partial\tenergy}{\partial\para},  \Wb}
        &:= \lim_{\eps\rightarrow 0 } \frac{\tenergy[\para + \eps\Wb, (q^1[\para])^{*_{\eps\Wb}},  (q^2[\para])^{*_{\eps\Wb}}] - \tenergy[\para,q^1[\para],q^2[\para]]}{\eps}\formPeriod
\end{align*}
In praxis this means to differentiate everything of the energy \wrt\ $ \para $ except contravariant proxy components of $ \qb[\para] $,
\eg\ in the term $ \int_{\surf} \|\qb[\para]\|^2 \mu[\para] = \int_{\surf} g_{ij}[\para] q^i[\para] q^j[\para] \mu[\para] $ has only $ g_{ij}[\para] $ and $ \mu[\para] $ to be differentiated.
Moreover, comparing with \eqref{eq:deformation_transport_vector} reveals 
\begin{align*}
    \innerH{\tangentR}{\frac{\partial\tenergy}{\partial\para},  \Wb}
        &= \innerH{\tangentR}{\frac{\partial\energy}{\partial\para},  \Wb}
            + \innerH[4]{\tangentS}{\frac{\del\energy}{\del \qb}, \gradW \qb[\para] } \formComma\\
    \innerH[4]{\tangentS[^0]}{\frac{\del\tenergy}{\del q^i}, \gauge{\Wb}q^i[\para] }
        &= \innerH[4]{\tangentS}{\frac{\del\energy}{\del \qb}, \gauge[\sharp]{\Wb} \qb[\para]}  \\
        &= \innerH[4]{\tangentS}{\frac{\del\energy}{\del \qb}, \gauge[\surf]{\Wb} \qb[\para] - \gradW\qb[\para]}\formPeriod
\end{align*}

This also affects the $ \hil $-gradient \eqref{eq:L2Grad_strong} in section \ref{sec:consequences_gradient_flow_vector}, 
which becomes
\begin{align}
    \hilgrad\tenergy
       &=  \left( \frac{\del\tenergy}{\del\para}, \left\{\frac{\del\tenergy}{\del q^i} \right\} \right)
       = \left( \frac{\partial\tenergy}{\partial\para} - \div_{\tangentS[^2]}\tgstress, \left\{\frac{\del\tenergy}{\del q^i} \right\} \right) \formComma
    \label{eq:L2Grad_strong_contravariant_approach}
\end{align}
where $ \tgstress = \gstress - \frac{\del\energy}{\del\qb}\otimes\qb[\para]\in\tangentS[^2] $ is given in table \ref{tab:gauge_stress} depending on the chosen gauge.

\subsection{Approach to \eqref{eq:deformation_transport_vector} with explicit covariant metric proxy} \label{sec:deformation_transport_vector_with_metric}
Some readers would like to include the covariant metric proxy 
$ \{g_{ij}[\para]\} = \{\inner{\tangentR}{\partial_i\para, \partial_j\para}\}\in(\tangentS[^0])^{2\times2} $,
so that the energy $ \energy $ can be rewritten as
\begin{align*}
    \henergy[\para, \qb[\para], \{g_{ij}[\para]\}] 
        &:= \energy[\para, \qb[\para]]
\end{align*}
without changing the sematic of $ \U $.
The total variation 
\begin{align*}
    \left\langle \frac{\del\henergy}{\del\para},  \Wb \right\rangle
            &:= \lim_{\eps\rightarrow 0 } \frac{\henergy[\para + \eps\Wb, \qb[\para + \eps\Wb], \{g_{ij}[\para] + \eps\Wb\}] 
                - \henergy[\para, \qb[\para], \{g_{ij}[\para]\}]}{\eps}
\end{align*}
\wrt\ $ \para $ in arbitrary directions of $ \Wb\in\tangentR $ equals $ \left\langle \frac{\del\energy}{\del\para},  \Wb \right\rangle $ \eqref{eq:variation_X_raw_vector}
as expected for a total derivative.
Also the total variation
\begin{align*}
     \left\langle \frac{\del\henergy}{\del\qb},  \rb \right\rangle
            &:= \lim_{\eps\rightarrow 0 } \frac{\henergy[\para, \qb[\para] + \eps\rb, \{g_{ij}[\para]\}] - \henergy[\para, \qb[\para], \{g_{ij}[\para]\}]}{\eps}
\end{align*}
\wrt\ $ \qb[\para] $ in arbitrary directions of $ \rb\in\tangentS $ equals $ \left\langle \frac{\del\energy}{\del\qb},  \rb \right\rangle $ \eqref{eq:variation_q_raw_vector}.
The Taylor expansion
\begin{align*}
    g_{ij}[\para+\eps\Wb]
        &= (g_{ij}[\para])^{*_{\eps\Wb}}
           + \eps (S_{ij}[\para,\Wb])^{*_{\eps\Wb}} +\landau(\eps^2)
\end{align*}
with scalar-valued pushforward $ *_{\eps\Wb}: \tangentS[^0] \rightarrow \tangent[^0]{\surf_{\eps\Wb}} $ \eqref{eq:pushforward_scalar}
yields
\begin{align*}
    \henergy[\para + \eps\Wb, \qb[\para + \eps\Wb], \{g_{ij}[\para+\eps\Wb]\}] \hspace{-120pt}\\
        &=  \henergy[\para + \eps\Wb, \proj_{\tangent{\surf_{\eps\Wb}}}\left(\qb[\para]\right)^{*_{\eps\Wb}}, \{(g_{ij}[\para])^{*_{\eps\Wb}}\}]\\
        &\quad\quad    + \eps \inner{\hilbertT{\surf_{\eps\Wb}}}{\frac{\del\henergy}{\del(\proj_{\tangent{\surf_{\eps\Wb}}}\qb^{*_{\eps\Wb}})} 
                                                        , \proj_{\tangent{\surf_{\eps\Wb}}}\left(\gauge{\Wb}\qb[\para]\right)^{*_{\eps\Wb}}}\\
        &\quad\quad    + \eps \innerH[4]{\tangent[^0]{\surf_{\eps\Wb}}}{ \frac{\partial\henergy}{\partial g_{ij}^{*_{\eps\Wb}}}, (S_{ij}[\para,\Wb])^{*_{\eps\Wb}} }
            + \landau(\eps^2)\formComma
\end{align*}
where  $*_{\eps\Wb}$ on vector fields applies $ \R^3 $-component-wise, see \eqref{eq:pushforward_vector_R3}.
For partial variations \wrt\ metric components see \eqref{eq:partial_variation_metric} below, but on $ \surf_{\eps\Wb} $ instead.
Since $*_{\eps\Wb} \rightarrow \Id $ on their respective spaces and  $\proj_{\tangent{\surf_{\eps\Wb}}} \rightarrow \proj_{\tangentS} $ for 
$ \eps\rightarrow0 $, we obtain
\begin{align}
    \inner[4]{\hilbertR}{\frac{\del\henergy}{\del\para} , \Wb}
        &= \inner[4]{\hilbertR}{\frac{\partial\henergy}{\partial\para} , \Wb}
            + \inner[4]{\hilbertS}{ \frac{\del\henergy}{\del\qb},  \gauge[\surf]{\Wb}\qb[\para] }  \notag\\ 
        &\quad\quad  + \innerH[4]{\tangentS[^0]}{ \frac{\partial\henergy}{\partial g_{ij}}, S_{ij}[\para,\Wb] } \formComma
     \label{eq:deformation_transport_vector_with_metric}
\end{align}
where the partial variations \wrt\ covariant metric proxy components in direction of scalar fields $ f\in\tangentS $ are defined by
\begin{align}
    \inner{}{ \frac{\partial\henergy}{\partial g_{ij}}, f }
            &:= \lim_{\eps\rightarrow 0 } \frac{\henergy[\para, \qb[\para], \{ g_{ij}[\para] + \eps f \}] - \henergy[\para, \qb[\para], \{ g_{ij}[\para] \}]}{\eps}
    \label{eq:partial_variation_metric}
\end{align}
and the partial variation \wrt $ \para $ in directions of $ \Wb\in\tangentR $ by
\begin{align*}
    \left\langle \frac{\partial\henergy}{\partial\para},  \Wb \right\rangle
            := \lim_{\eps\rightarrow 0 } 
                \frac{1}{\eps}\Big( &\henergy[\para + \eps\Wb, \proj_{\tangent{\surf_{\eps\Wb}}}\left(\qb[\para]\right)^{*_{\eps\Wb}}, \{(g_{ij}[\para])^{*_{\eps\Wb}}\}]\\ 
                                    & - \henergy[\para, \qb[\para], \{ g_{ij}[\para] \}]\Big) \formPeriod
\end{align*}
In praxis this means to differentiate everything of the energy \wrt\ $ \para $ except for $ \qb[\para] $ (incl.\ its frame) and explicit terms of $ g_{ij}[\para] $,
\eg\ in the term $ \int_{\surf} \|\qb[\para]\|^2 \mu[\para] = \int_{\surf} \|\qb[\para]\|^2 \widehat{\mu}[\{ g_{ij}[\para] \}] $ nothing has to be differentiated,
\ie\ $ \left\langle \frac{\partial\henergy}{\partial\para},  \Wb \right\rangle = 0 $.
Moreover, comparing with \eqref{eq:deformation_transport_vector} reveals 
\begin{align*}
     \inner[4]{\hilbertR}{\frac{\partial\henergy}{\partial\para} , \Wb}  
     + \innerH[4]{\tangentS[^0]}{ \frac{\partial\henergy}{\partial g_{ij}}, S_{ij}[\para,\Wb] }
        &= \inner[4]{\hilbertR}{\frac{\partial\energy}{\partial\para} , \Wb} 
\end{align*}
and $ \inner[2]{\hilbertS}{ \frac{\del\henergy}{\del\qb},  \gauge[\surf]{\Wb}\qb[\para] } = \inner[2]{\hilbertS}{ \frac{\del\energy}{\del\qb},  \gauge[\surf]{\Wb}\qb[\para] } $
as expected for a total derivative \wrt\ same arguments.
Note that we could substitute 
$ \innerH[2]{\tangentS[^0]}{ \frac{\partial\henergy}{\partial g_{ij}}, S_{ij}[\para,\Wb] } = \innerH[2]{\tangentS[^0]}{ \frac{\partial\henergy}{\partial g_{ij}}, G_{ij}[\para,\Wb] }$
due to symmetry.

\section{Outsourced Calculations}

\subsection{Deformation Derivative on Geometrical Quantities}

\subsubsection{Metric Tensor, its Inverse and Density}\label{sec:deformation_of_metric_density}
With the metric tensor proxy, given by $ g_{ij}[\para] = \inner{\tangentR}{\partial_i\para, \partial_j\para} $, 
contravariant proxy components $ G_{ij}[\para,\Wb] $ \eqref{eq:tangential_part_of_gradW}  of the tangential derivative 
of deformation direction $ \Wb\in\tangentR $ 
and the symmetric part $S_{ij}[\para,\Wb]$, the deformation derivative \eqref{eq:deformation_derivativ_scalar} on scalar fields yields
\begin{align}
    \gauge{\Wb}g_{ij}[\para]
        &= \inner{\tangentR}{\partial_i\Wb, \partial_j\para} + \inner{\tangentR}{\partial_i\para, \partial_j\Wb}
         = G_{ij}[\para,\Wb] + G_{ji}[\para,\Wb]\notag\\
        &= 2 S_{ij}[\para,\Wb] \formPeriod
    \label{eq:deformation_of_metric}
\end{align}
Due to this, the inverse tensor proxy gives
\begin{align}
    \gauge{\Wb}g^{ij}[\para]
        &=  \gauge{\Wb}\left( g^{ik}[\para]g^{jl}[\para]g_{kl}[\para] \right)
         = 2 \gauge{\Wb}g^{ij}[\para] + 2 S^{ij}[\para,\Wb]\notag\\
        &= - 2 S^{ij}[\para,\Wb]
    \label{eq:deformation_of_inverse_metric}
\end{align}
and the density value $ \sqrt{\vert\gb\vert}[\para] := \sqrt{\det\qb[\para]}\in\tangentS[^0] $ yields
\begin{align*}
    \gauge{\Wb} \sqrt{\vert\gb\vert}[\para]
        &= \sqrt{\vert\gb\vert}[\para]\Tr\Gb[\para,\Wb] \formComma
\end{align*}
\resp\ $ \gauge{\Wb}\mu[\para] = \Tr\Gb[\para,\Wb]\mu[\para] $,
where $ \mu[\para] = \sqrt{\vert\gb\vert}[\para]dy^1\!\wedge\! dy^2 \cong \Eb[\para] $.
It is noteworthy that both the Jaumann and material deformation derivative are metric-compatible, \ie\ the tangential field identity map $[\Id_{\tangentS}]_{ij}=g_{ij}$ yields
\begin{align*}
    \gaugeS[\Wb]\Id_{\tangentS} &=  \gaugeJ[\Wb]\Id_{\tangentS} = \nullb \formPeriod
\end{align*}
This follows from $[\gaugeLL[\Wb]\Id_{\tangentS}]_{ij} = \gauge{\Wb}g_{ij}[\para] $, $[\gaugeUU[\Wb]\Id_{\tangentS}]^{ij} = \gauge{\Wb}g^{ij}[\para] $, \eqref{eq:jaumann_deformation_derivativ_relation_to_convected_tensor} and \eqref{eq:jaumann_deformation_derivativ_tensor}.
As a consequence we get trace-compatibility for all $\qb[\para]\in\tangentS[^2]$:
\begin{align}\label{eq:trace_compatibility}
    \gauge{\Wb} \Tr\qb[\para] 
        &= \gauge{\Wb} \inner{\tangentS[^2]}{\qb[\para], \Id_{\tangentS}}
         = \Tr \gaugeS[\Wb] \qb[\para] 
         = \Tr \gaugeJ[\Wb] \qb[\para] \formPeriod
\end{align}

\subsubsection{Christoffel Symbols}
With Christoffel symbols of first kind \eqref{eq:Christoffel_first_kind} and contravariant proxy components $ G_{ij}[\para,\Wb] $ \eqref{eq:tangential_part_of_gradW}  of the tangential derivative 
of deformation direction $ \Wb\in\tangentR $, the deformation derivative \eqref{eq:deformation_derivativ_scalar} on scalar fields yields
\begin{align*}
    \gauge{\Wb}\Gamma_{jkl}[\para]
        &= \inner{\tangentR}{\partial_j\partial_k\Wb, \partial_l\para} 
            + \inner{\tangentR}{\partial_j\partial_k\para, \partial_l\Wb}\\
        &= \partial_j G_{lk}[\para,\Wb] 
            - \inner{\tangentR}{\partial_j\partial_l\para, \partial_k\Wb}
            + \inner{\tangentR}{\partial_j\partial_k\para, \partial_l\Wb}
\end{align*}
The full surface gradient $ \nabla_{\surf}\Wb = g^{il} (\partial_l\Wb) \otimes \partial_i\para\in\tangentR\otimes\tangentS $ is given in \eqref{eq:surface_derivative_of_W}, 
\resp\ \eqref{eq:tangential_derivative_of_W_components} in components.
The covariant Hessian components of $ \para $ reads
$ \partial_j\partial_k\para = \Gamma_{jk}^m[\para]\partial_m\para + \shopC_{jk}[\para]\normal[\para] $.
Therefore, we obtain
\begin{align*}
    \inner{\tangentR}{\partial_j\partial_k\para, \partial_l\Wb}
        &= \Gamma_{jk}^m[\para]G_{ml}[\para,\Wb] + b_{l}[\para,\Wb]\shopC_{jk}[\para]\formComma
\end{align*}
where $ \bb[\para,\Wb]=\nabla\wnor[\para] + \shop[\para]\wb[\para]\in\tangentS $ determines the normal-tangential part of $ \nabla_{\surf}\Wb$, see \eqref{eq:b_definition}.
Summing all things up and using the deformation of the inverse metric in \eqref{eq:deformation_of_inverse_metric}, the deformation derivative of the Christoffel symbols of second kind yields
\begin{align}
    \gauge{\Wb}\Gamma_{jk}^i[\para]
        &= \gauge{\Wb}\left( g^{il}[\para] \Gamma_{jkl}[\para] \right)
         = g^{il} \gauge{\Wb}\Gamma_{jkl}[\para] - 2 S^{il}[\para,\Wb]\Gamma_{jkl}[\para] \notag\\
        &= \tensor{G}{^i_{k\vert j}}[\para,\Wb] 
                + b^i[\para,\Wb]\shopC_{jk}[\para] - b_k[\para,\Wb]\shopC^i_j[\para]\formPeriod
   \label{eq:deformation_of_christoffel_second}
\end{align}

\subsubsection{Normals Field}
The deformation derivative $ \gauge{\Wb}\normal[\para] = (\gauge{\Wb}\normalC^{I}[\para])\eb_{I}\in\tangentS $ is tangential, since $ \gauge{\Wb} $ obeys the product rule 
\wrt\ the local inner product and the normals field $ \normal[\para]\in\tangentR $ is normalized.
Therefore, with \eqref{eq:normal_gradient_of_W} and \eqref{eq:b_definition}, we can simply calculate
\begin{align}
    \gauge{\Wb}\normal[\para]
        &= g^{ij}[\para]\inner{\tangentR}{\gauge{\Wb}\normal[\para], \partial_j\para}\partial_i\para
         = -g^{ij}[\para]\inner{\tangentR}{\normal[\para], \partial_j\Wb}\partial_i\para\notag\\
        &= -\left( \nabla\wnor[\para] + \shop[\para]\wb[\para] \right)
         = -\bb[\para,\Wb] \formPeriod
    \label{eq:deformation_of_normals}
\end{align}

\subsubsection{Second Fundamental Form}
With the deformation derivative \eqref{eq:deformation_of_normals} of the normal field  and the normal part of the gradient of $ \Wb $ in \eqref{eq:normal_gradient_of_W},
the deformation derivative \eqref{eq:deformation_derivativ_scalar} of the covariant proxy components \eqref{eq:shape_operator} of the second fundamental form yields
\begin{align*}
    \gauge{\Wb}\shopC_{ij}[\para]
        &= \inner{\tangentR}{\partial_j\partial_i\Wb, \normal[\para]} - \inner{\tangentR}{\partial_j\partial_i\para, \bb[\para,\Wb]}\\
        &= \partial_j \inner{\tangentR}{\partial_i\Wb, \normal[\para]} 
                +\inner{\tangentR}{\partial_i\Wb, \shopC^k_j[\para]\partial_k\para}
                - \Gamma_{ij}^k[\para]b_k [\para,\Wb] \\
        &= b_{i\vert j}[\para,\Wb] + G_{ki}[\para,\Wb]\shopC^k_j[\para] 
          = \left[ \nabla\bb[\para,\Wb] + \Gb^{T}[\para,\Wb]\shop[\para] \right]_{ij}\formPeriod
\end{align*}
Since $ \left[ \gaugeLL[\Wb]\shop[\para] \right]_{ij} = \gauge{\Wb}\shopC_{ij}[\para] $ is valid by \eqref{eq:deformation_derivative_components_natural_tensor},
the relation \eqref{eq:deformation_derivative_natural_tensor} to the material deformation derivative \eqref{eq:deformation_derivativ_tensor} results in
\begin{align}
    \gaugeS[\Wb]\shop[\para]
     &= \nabla\bb[\para,\Wb] - \shop[\para]\Gb[\para,\Wb] \formPeriod
   \label{eq:deformation_of_shape_operator}
\end{align}
As a consequence of trace-compatibility \eqref{eq:trace_compatibility}, the deformation derivative of the mean curvature $\meanc[\para] = \Tr\shop[\para]$ yields
\begin{align}
    \gauge{\Wb}\meanc[\para]
        &= \div\bb[\para,\Wb] - \inner{\tangentS[^2]}{\shop[\para],\Gb[\para,\Wb]} \formPeriod
    \label{eq:deformation_of_mean_curvature}
\end{align}

\subsection{Calculations for Example \ref{sec:example_vector}}\label{sec:calculation_for_example}

The deformation derivative \eqref{eq:deformation_of_christoffel_second} of Christoffel symbols and the upper-lower deformation derivative 
    in \eqref{eq:deformation_derivative_components_natural_tensor} yields
    \begin{align*}
        \tensor{\left[ \gaugeUL[\Wb]\nabla\qb[\para]\big\vert_{\gaugeU\qb=0} \right]}{^i_k}
            &= \gauge{\Wb} \tensor{q}{^i_{\vert k}}[\para]\big\vert_{\gaugeU\qb=0}
             = q^j[\para] \gauge{\Wb} \Gamma_{kj}^{i}[\para]\\
            &= \big[ \nabla_{\qb[\para]}\Gb[\para,\Wb] 
                            - \shop[\para]\qb[\para]\otimes\bb[\para,\Wb]\\
             &\hspace{85pt} +\bb[\para,\Wb]\otimes\shop[\para]\qb[\para]\tensor{\big]}{^i_k} \formComma
    \end{align*}
    since $ \gauge{\Wb}\partial_k q^{i}[\para] = 0 $ is valid for $\gauge{} q^{i}=0$.
    Using the relation \eqref{eq:deformation_derivative_natural_tensor} between the material and the upper-lower deformation derivative and the identity
    \begin{align*}
        \nabla_{\qb[\para]}\Gb[\para,\Wb] - \shop[\para]\qb[\para]\otimes\bb[\para,\Wb] \hspace{-80pt}\\
            &= \qb[\para]\nabla\Gb^T[\para,\Wb] -\inner{\tangentS}{\qb[\para],\bb[\para,\Wb]}\shop[\para] \formComma
    \end{align*}
    we can relate this to the material deformation derivative \eqref{eq:deformation_derivativ_tensor} by
    \begin{align*}
        \gaugeS[\Wb]\nabla\qb[\para]\big\vert_{\gaugeU\qb=0}
            &= \qb[\para]\nabla\Gb^T[\para,\Wb] + \Gb[\para,\Wb]\nabla\qb[\para] - (\nabla\qb[\para])\Gb[\para,\Wb] \\
            &\quad -\inner{\tangentS}{\qb[\para],\bb[\para,\Wb]}\shop[\para] +\bb[\para,\Wb]\otimes\shop[\para]\qb[\para]\formPeriod
    \end{align*}
    Compatibility of the material deformation derivative $ \gaugeS[\Wb] $ \wrt\ the inner product and 
    density representation \eqref{eq:deformation_transport_theorem} of the energy $\energy_{\nabla}$  yields 
    \begin{align*}
        \inner[4]{\hilbertR}{\frac{\del\energy_{\nabla}}{\del\para}\Big\vert_{\gaugeU\qb=0} , \Wb} \hspace{-25pt}
                   &= K\Big( \inner[2]{\hilbertS[^2]}{ \gaugeS[\Wb]\nabla\qb[\para]\big\vert_{\gaugeU\qb=0}, \nabla\qb[\para]  } \\
                    &\quad +\frac{1}{2}\inner[2]{\hilbertS[^2]}{\left\| \nabla\qb[\para] \right\|^2_{\tangentS[^2]}\Id_{\tangentS}, \Gb[\para,\Wb]}\Big)\\
              &= K \Big( \inner{\hilbertS[^3]}{ \qb[\para]\otimes\nabla\qb[\para], \nabla\Gb^T[\para,\Wb] }\\
                   &\quad + \inner{\hilbertS[^2]}{ (\nabla\qb[\para])(\nabla\qb[\para])^T - \sigmab_{E,I}, \Gb[\para,\Wb] }\\
                   &\quad +  \inner{\hilbertS}{ \nabla_{(\shop[\para]\qb[\para])}\qb[\para] - (\nabla_{\shop[\para]}\qb[\para])\qb[\para], \bb[\para,\Wb] }
                  \Big)
    \end{align*}
    with $\frac{\sigmab_{E,I}}{K}
                    = \proj_{\mathcal{Q}^2\surf}( (\nabla\qb[\para])^T \nabla\qb[\para] )
                     = (\nabla\qb[\para])^T \nabla\qb[\para] - \frac{\left\| \nabla\qb[\para] \right\|^2_{\tangentS[^2]}}{2} \Id_{\tangentS}$.
    Eventually, integration by parts and the identity 
    \begin{align*}
        \left( \div(\qb[\para]\otimes\nabla\qb[\para]) \right)^T
            &= (\nabla\qb[\para])(\nabla\qb[\para])^T + (\Delta\qb[\para])\otimes\qb[\para]
    \end{align*}
    result in \eqref{eq:partial_variation_UNabla}.

\section*{Declarations}

\subsection*{Ethical approval} 
not applicable
 
\subsection*{Competing interests} 
There are no competing interests. 
 
\subsection*{Authors' contributions} 
This project was conceived by I.N., S.S., A.V. and I.N. derived the theory and performed the
simulations. The results were analysed by I.N., S.S. and A.V. The main text was written by 
I.N. and A.V.

\subsection*{Funding} 
A.V. was supported by DFG through FOR3013.
S.S. is grateful for the support he received from TU Dresden in the form of a Dresden Junior Fellowship which allowed for this collaboration.
 
\subsection*{Availability of data and materials} 
The datasets generated and/or analysed during the current study are available from the corresponding author on reasonable request.

\bibliography{goi_bib}
\bibliographystyle{abbrvnat}

\end{document}